\documentclass[twocolumn,aps,superscriptaddress,floatfix]{revtex4}
\usepackage{amsmath}
\usepackage{amssymb}
\usepackage{graphicx}
\usepackage{color}

\newcommand{\be}{\begin{equation}}
\newcommand{\ee}{\end{equation}}
\newcommand{\bea}{\begin{eqnarray}}
\newcommand{\eea}{\end{eqnarray}}
\newcommand{\bse}{\begin{subequations}}
\newcommand{\ese}{\end{subequations}}
\newcommand{\eca}{EuCo$_{2-y}$As$_2$}

\begin{document}

\title{EPR measurements of Eu$^{+2}$ spins in metallic EuCo$_{2-y}$As$_2$ single crystals}

\author{N. S. Sangeetha}
\affiliation{Ames Laboratory, Iowa State University, Ames, Iowa 50011, USA}
\author{S. D. Cady}
\affiliation{Chemical Instrumentation Facility, Iowa State University, Ames, Iowa 50011, USA}
\author{D. C. Johnston}
\affiliation{Ames Laboratory, Iowa State University, Ames, Iowa 50011, USA}
\affiliation{Department of Physics and Astronomy, Iowa State University, Ames, Iowa 50011, USA}

\date{\today}

\begin{abstract}

The Eu$^{+2}$ spins~$S=7/2$ in the metallic compound \eca\ order into an antiferromagnetic helical structure below a N\'eel temperature $T_{\rm N} = 40$--45~K\@.  The effective magnetic moment $\mu_{\rm eff}$ of the Eu spins in the paramagnetic state from 100 to 300~K is found from static magnetic susceptibility measurements to be enhanced by about 7\% compared to the value expected for spectroscopic splitting factor $g=2$, and the saturation moment at high applied fields~$H$ and low temperatures~$T$ is also sometimes enhanced.  Here electron-paramagnetic-resonance (CW~EPR) measurements versus applied magnetic field~$H$ at fixed X-band rf (microwave) angular frequency~$\omega$  were carried out using a linearly-polarized rf magnetic field oriented perpendicular to~{\bf H} to study the microscopic magnetic properties of the Eu spins.  In order to analyze the data, the complex magnetic susceptibility $\chi(\omega)$ at fixed~$H$ was used that was derived for linearly-polarized rf fields from the modified Bloch equations~[M. A. Garstens and J. I. Kaplan, Phys. Rev. {\bf 99}, 459 (1955)] (GK).  The validity of their $\chi(\omega)$ was verified by showing that the dispersive part can be derived from the absorptive part using a Kramers-Kronig relation. It is shown that  their formulation when applied to calculate the Dysonian absorptive susceptibility $\chi_{\rm D}^{\prime\prime}(H)$ of local magnetic moments in metals yields a prediction that can be very different from the traditionally-used form of $\chi_{\rm D}^{\prime\prime}(H)$.  By fitting the derivative of the field-swept CW~EPR data for \eca\ by $\chi_{\rm D}^{\prime\prime}(H)$ at fixed~$\omega$ derived from the GK $\chi_{\rm D}^{\prime\prime}(\omega)$ at fixed~$H$, the Eu spin spectroscopic splitting factor ($g$-factor) is found to be $\approx 2.00$ from 300 to $\sim125$~K, and then to continuously increase to $\approx 2.16$ on further cooling to 50~K\@.  We speculate that the enhancement of the Eu $g$-factor on cooling from $\sim 125$ to~50~K  arises from continuously-increasing local short-range ferromagnetic correlations between the Co $3d$-band electrons and the Eu spins.

\end{abstract}

\maketitle

\section{Introduction}

The $s$-state ions Eu$^{+2}$ and Gd$^{+3}$ have electronic configuration $4f^7$ with spin $S=7/2$ and orbital angular momentum $L=0$.  Therefore spin-orbit coupling of these ions to the lattice is very weak and their spectroscopic splitting factors ($g$-factors) are usually close to the free-electron value of 2.  The paramagnetic~(PM) effective moment for $g=2$ and $S=7/2$ is $\mu_{\rm eff} = g\sqrt{S(S+1}\,\mu_{\rm B} = 7.94~\,\mu_{\rm B}$, where $\mu_{\rm B}$ is the Bohr magneton.  In most compounds containing these ions, static magnetic susceptibility $\chi$ versus temperature~$T$  measurements in the PM state reveal $\mu_{\rm eff}$ values close to this value.  For example, for the Eu$^{+2}$ spins in the PM state of the helical antiferromagnet ${\rm EuCo_2P_2}$ with the body-centered-tetragonal ${\rm ThCr_2Si_2}$ structure, to within the errors $\mu_{\rm eff}$ is equal to the predicted value and the ordered (saturation) moment at low temperatures is also equal to the predicted value $\mu_{\rm sat} = gS\,\mu_{\rm B} = 7\,\mu_{\rm B}$/Eu atom \cite{Sangeetha2016}.  

However, in the isostructural helical antiferromagnet ${\rm EuCo_2As_2}$, $\mu_{\rm eff}$ over the $T$ range from 100 to 300~K was found to be about $8.5~\mu_{\rm B}$/Eu~atom, corresponding to a 7\% enhancement of the $g$-factor~\cite{Sangeetha2017}.  From electronic structure calculations, this enhancement was deduced to be due to ferromagnetic (FM) polarization of the Co $3d$ electrons~\cite{Sangeetha2017} in the static magnetic field~$H$ applied during the $\chi(T)$ measurements.  One can envision two scenarios for the effect this polarization might have on the $g$-factor of the Eu spins.  In one scenario, the polarization of the Co 3$d$ electrons would have no influence on the $g$-factor of the Eu spins due to lack of correlations between them.  Alternatively, an enhanced $g$-factor could originate from FM correlations between the itinerant Co $3d$ band electrons and the Eu spins.  In the latter case, one would expect microscopic measurements of the Eu spin $g$-factor to show an enhancement of about 7\%, whereas in the former case not so much.  The present Eu electron-paramagnetic-resonance (CW~EPR) measurements were carried out in the PM state of \eca\ from 50~K to 180~K as a microscopic probe of the degree to which the Eu $g$-factor is enhanced, if at all. In the $T$ region between $\sim 125$ and 180~K over which the $\chi(T)$ masurements found the enhanced $\mu_{\rm eff}$, the Eu spin spectroscopic-splitting factor ($g$-factor) is found to be nearly $T$ independent with an unenhanced $g\approx 2.00$. On the other hand, when cooling further to 50~K, the $g$-factor increases by about 8\% to $\approx 2.16$.  We speculate that the unenhanced $g$-factor from  $\sim 125$ to~180~K occurs due to negligible local FM correlations between the Eu spins and the conduction-electron spins, but that with further cooling such short-range FM local correlations increasingly develop that result in an increase in the $g$-factor of the Eu spins.

Because \eca\ is metallic, we expected and found a Dysonian lineshape~\cite{Dyson1955, Feher1955} in our CW~EPR spectra measured versus $H$ at fixed rf angular frequency~$\omega$.  Due to the large linewidths observed, one must derive the complex frequency-dependent magnetic susceptibility
\be
\chi(\omega) = \chi^\prime(\omega) - i\chi^{\prime\prime}(\omega)
\label{Eq:chippp(omega)}
\ee
from the modified Bloch equations \cite{Garstens1955, Abragam1961} instead of the Bloch equations~\cite{Bloch1946}.  The difference is that the relaxation of the local-moment magnetization in the Bloch equations is towards the static applied magnetic field~{\bf H}, whereas in the modified Bloch equations the relaxation is towards the instantaneous magnetic field, which includes both {\bf H} and rf magnetic fields, where the polarization of the latter is perpendicular to {\bf H}.  The modified Bloch equations are used in order that $\chi(\omega)$ gives physically correct limits as discussed later.  Furthermore, CW EPR spectrometers are typically operated using a linearly-polarized rf magnetic field and that feature must be taken into account when calculating $\chi(\omega)$~\cite{Garstens1955}. 

In Sec.~\ref{Sec:ExpDet} the experimental details are given.  The theory needed to analyze our EPR data is given in Sec.~\ref{Sec:Theory}.  In order to contrast $\chi(\omega)$ obtained from the Bloch equations from that obtained from the modified Bloch equations, the former is discussed in Sec.~\ref{Sec:Bloch}.  The modified Bloch equations and the solution of $\chi(\omega)$ at fixed $H$  obtained from them~\cite{Garstens1955} and the experimentally-relevant $\chi(H)$ at fixed $\omega$ are given in Sec.~\ref{Sec:ModifiedBloch}.  Expressions for the power absorbed by a sample from the rf magnetic field and the related skin depth are presented for both scanning $\omega$ at fixed~$H$ and scanning~$H$ at fixed~$\omega$ in Sec.~\ref{Sec:PowerSkinDepth}.  The Dysonian absorptive susceptibility $\chi_{\rm D}^{\prime\prime}(\omega)$ is discussed in Sec.~\ref{Eq:DysonEPR}, where~\cite{Dyson1955}
\bse
\label{Eq:chiDpp}
\be
\chi_{\rm D}^{\prime\prime}(\omega) = \chi^{\prime\prime}(\omega) +\alpha\chi^\prime(\omega)
\ee
and the Dysonian lineshape parameter $\alpha$ has the range
\be
0\leq\alpha\leq1.
\ee
\ese
A comparison of $\chi_{\rm D}^{\prime\prime}(\omega)$ obtained from the modified Bloch equations with a previously-used expression is given in Sec.~\ref{Sec:PreviousLinshapes}, where the latter is shown to become quite different from the former with increasing~$\alpha$ and increasing linewidth.

The results of our CW~EPR measurements on \eca\ and their analyses in terms of the predictions obtained from the modified Bloch equations in Sec.~\ref{Sec:Theory} are given in Sec.~\ref{Sec:Results}.   An overview of the spectra is given in Sec.~\ref{Sec:EPR Overview}.  The temperature dependences of the fitted $\alpha$ parameter, resonance field $H_{\rm res}$ and internal resonance field $H_{\rm res}^{\rm int}$, $g$-factor, and Lorentzian line half-width $\Delta H$ are presented in the remainder of Sec.~\ref{Sec:Results}.  A summary and discussion are given in Sec.~\ref{Sec:Summary}. 

\section{\label{Sec:ExpDet} Experimental Details}

Two single crystals of \eca\ labeled Crystals~\#2 and~\#3 were studied that were taken from the same batches of crystals from which extensive crystallograhic and physical-property data for crystals also labeled Crystals~\#2 (Sn-flux-grown) and~\#3 (CoAs self-flux-grown) were presented in Ref.~\cite{Sangeetha2017}.

The CW EPR measurements were carried out at a fixed X-band frequency of 9.380~GHz and magnetic field scan range 0 to 6~kOe using an Elexsys E580 FT/CW EPR spectrometer in CW mode.  The static magnetic field~{\bf H} was applied along the $c$~axis of the two crystals measured.  The microwave magnetic field was perpendicular to {\bf H} and hence directed along the two large flat surfaces of the crystals parallel to the $ab$~plane.  The EPR data reported here cover the $T$~range from 50~K to 180~K which is in the PM temperature region of the two crystals above their respective N\'eel temperatures of 45~K for Crystal\#2 and 40~K for Crystal~\#3~\cite{Sangeetha2017}. To improve the signal-to-noise ratio, the applied magnetic field was modulated at a frequency of 100~kHz with lock-in amplifier detection at that frequency, so a measured EPR spectrum is the field derivative of $\chi_{\rm D}^{\prime\prime}(H)$ (see Sec.~\ref{Sec:EPR Overview} below).

The Gaussian cgs system of units is used in this paper, with the exception of the expression for the skin depth $\delta$ which is expressed in SI units.

\section{\label{Sec:Theory} Theory}

\subsection{\label{Sec:Bloch} Bloch Equations}

In both nuclear (NMR) \cite{Abragam1961, Bloch1946, Slichter1963} and electron (EPR) \cite{Abragam1970, Pake1973, Taylor1975, Barnes1981, Poole1983} magnetic resonance and relaxation, the Bloch equations are often the starting point for analyzing experimental data if the resonances are sufficiently narrow. They give the Cartesian components of the magnetization~{\bf M} (average magnetic moment per unit volume), which is precessing around the magnetic field {\bf H}, as \cite{Bloch1946}
\bse
\label{Eqs:Bloch}
\bea
\frac{dM_x}{dt} &=& \gamma ({\bf M}\times{\bf H})_x - M_x/T_2,\\
\frac{dM_y}{dt} &=& \gamma ({\bf M}\times{\bf H})_y - M_y/T_2,\\
\frac{dM_z}{dt} &=& \gamma ({\bf M}\times{\bf H})_z + (M_0-M_z)/T_1,
\eea
\ese
where $t$ is the time, $M_0 = \chi_0 H_0$ is the thermal-average magnetization per unit volume  when the magnetization is aligned in the direction of the applied field
\be
{\bf H}_0 = H_0 \hat{\bf k},
\label{Eq:H0vec}
\ee
where we switch notation from the above {\bf H} to ${\bf H}_0$ since {\bf H} now contains the contribution from the rf magnetic field ${\bf H}_1$ in Eqs.~(\ref{Eqs:H1}) below.  Here $\chi_0$ is the dimensioneless static magnetic susceptibility per unit volume, $T_1$ is the longitudinal relaxation time associated with decay of the magnetic energy, $T_2$ is the transverse relaxation time associated with spin-spin interactions, and $\gamma$ is the gyromagnetic ratio of the moment ($\gamma = -g\mu_{\rm B}/\hbar$ for electronic Heisenberg spins, $g$ is the spectroscopic splitting factor also called the $g$-factor, $\mu_{\rm B}$ is the Bohr magneton, and $\hbar$ is Planck's constant~$h$ divided by $2\pi$).  The damping terms on the far right sides of Eqs.~(\ref{Eqs:Bloch}) are phenomenologically introduced so that the relaxation of each of the Cartesian components of {\bf M} in a free-induction decay experiment is exponential.

For magnetic resonance experiments, an additional radio-frequency (rf) (or microwave) magnetic field ${\bf H}_1$ with angular frequency $\omega$ is applied perpendicular to ${\bf H}_0$ that induces transitions between the quantum Zeeman energy levels and is taken here to be linearly polarized, as in most CW EPR experiments, which is assigned to be along the $x$~axis.  ${\bf H}_1$  can be considered to be a superposition of a circularly-polarized magnetic field that has a precession angular velocity parallel to the $z$~axis as does {\bf M} and a counter-rotating field that has a precession angular velocity antiparallel to the $z$~axis, i.e.,
\bse
\label{Eqs:H1}
\bea
{\bf H}_1 &=&  H_1 \cos(\omega t)\hat{\bf i} \label{Eq:LP} \\
	&=& \frac{H_1}{2}[\cos(\omega t)\hat{\bf i} + \sin(\omega t)\hat{\bf j}] \label{Eq:RC}\\
&& +\ \frac{H_1}{2}[\cos(\omega t)\hat{\bf i} - \sin(\omega t)\hat{\bf j}].\label{Eq:CRC}
\eea
\ese
For narrow resonance lines, the counter-rotating component of ${\bf H}_1$ in Eq.~(\ref{Eq:CRC}) makes no significant contribution to the observed EPR signal and is therefore usually ignored.  However, for wide EPR spectra with widths of the order of $H_0$, the influence of the counter-rotating component~(\ref{Eq:CRC}) of ${\bf H}_1$ must also be taken into account.

The response of $M_x$ to first order in $H_1$ (nonsaturating condition) is written
\be
M_x = H_1[\chi^\prime(\omega) \cos(\omega t) + \chi^{\prime\prime}(\omega)\sin(\omega t)],
\label{Eq:Mxchi}
\ee
where the complex magnetic susceptibility versus frequency $\chi(\omega)$ is given in Eq.~(\ref{Eq:chippp(omega)}).  The dispersive [$\chi^{\prime} (\omega)$] and absorptive [$\chi^{\prime\prime} (\omega)$] components of the steady-state $\chi(\omega)$ to first order in $H_1$ (no saturation) are obtained from Eqs.~(\ref{Eqs:Bloch}) using the method of Ref.~\cite{Garstens1955} for linearly-polarized rf fields that does not employ the technique of rotating reference frames as
\bea
\frac{\chi^{\prime}(\omega)}{\chi_0} &=& \frac{\omega_0T_2}{2}\bigg[\frac{(\omega+\omega_0)T_2}{(\omega+\omega_0)^2T_2^2+1} - \frac{(\omega-\omega_0)T_2}{(\omega-\omega_0)^2T_2^2+1} \bigg],\nonumber\\
\label{chipp10}\\
\frac{\chi^{\prime\prime}(\omega) }{\chi_0} &=& \frac{\omega_0T_2}{2}\bigg[\frac{1}{(\omega-\omega_0)^2T_2^2+1} - \frac{1}{(\omega+\omega_0)^2T_2^2+1} \bigg].\nonumber
\eea
These expressions contain the contributions of both the rotating and counter-rotating components of ${\bf H}_1$ in Eqs.~(\ref{Eqs:H1}). One sees that $\chi^{\prime}(\omega)$ is even in~$\omega$ and $\chi^{\prime\prime}(\omega)$ is odd in~$\omega$ as required. The Lorentzian half-width at half-maximum peak height~$\Delta\omega$ is related to $T_2$ by
\be
T_2 = 1/\Delta\omega.
\label{Eq:T2Dw}
\ee
Making this substitution into Eqs.~(\ref{chipp10}) gives
\bea
\frac{\chi^{\prime}(\omega)}{\chi_0} &=& \frac{\omega_0}{2} \bigg[\frac{\omega+\omega_0}{(\omega+\omega_0)^2+\Delta\omega^2} - \frac{\omega-\omega_0}{(\omega-\omega_0)^2+\Delta\omega^2} \bigg],\nonumber\\
\label{chipp20}\\
\frac{\chi^{\prime\prime}(\omega) }{\chi_0} &=& \frac{\omega_0\Delta\omega}{2}\bigg[\frac{1}{(\omega-\omega_0)^2+\Delta\omega^2} - \frac{1}{(\omega+\omega_0)^2+\Delta\omega^2} \bigg].\nonumber
\eea
However, there is a serious problem with the second of each of Eqs.~(\ref{chipp10}) and~(\ref{chipp20}).  As discussed later in Sec.~\ref{Sec:PowerSkinDepth}, the time-averaged rf power $P$ absorbed by a sample is proportional to $\omega\chi^{\prime\prime}(\omega)$. Thus the second of Eqs.~(\ref{chipp20}) predicts that $P$ is zero if $H_0=\omega_0/\gamma = 0$ even when $\omega$ is nonzero, which is unphysical~\cite{Abragam1961}. The modified Bloch equations discussed in the following section correct this error. 

\subsection{\label{Sec:ModifiedBloch} Modified Bloch Equations}

Instead of the magnetization {\bf M} relaxing towards the static magnetic field ${\bf H}_0$ as in the Bloch equations, in the modified Bloch equations {\bf M} relaxes towards the instantaneous magnetic field~{\bf H}, which from Eqs.~(\ref{Eq:H0vec}) and~(\ref{Eq:LP}) is
\be
{\bf H} = {\bf H}_1 + {\bf H}_0  =  H_1 \cos(\omega t)\hat{\bf i} + H_0 \hat{\bf k}.
\ee
Furthermore, there is no longer a distinction between $T_1$ and $T_2$ \cite{Abragam1961}, so the relaxation time is written as $\tau = T_1 = T_2$. The modified Bloch equations are then~\cite{Garstens1955}
\bse
\label{Eqs:ModBloch}
\bea
\frac{dM_x(t)}{dt} &=& \omega_0 M_y(t) - \frac{M_x(t)-\chi_0H_1\cos(\omega t)}{\tau},\\
\frac{dM_y(t)}{dt} &=& \gamma H_1 \cos(\omega t)M_z(t) - \omega_0 M_x(t) - \frac{M_y(t)}{\tau},\hspace{0.3in}\\
\frac{dM_z(t)}{dt} &=& -\gamma H_1 \cos(\omega t)M_y(t)- \frac{M_z(t)-\chi_0H_0}{\tau}.
\eea
\ese

Garstens and Kaplan~\cite{Garstens1955} obtained in 1955 a solution for ${\bf M}(t)$ from these equations for linearly-polarized rf fields that has the important feature that it automatically takes into account both the rotating and counter-rotating components of the rf magnetic field ${\bf H}_1$ in Eqs.~(\ref{Eqs:H1}), as already seen above in the solution to the Bloch equations in Eqs.~(\ref{chipp10}).  Proceeding as described in Ref.~\cite{Garstens1955}, to first order in $H_1$ (no saturation) the solution for  $M_x(t)$ in Eqs.~(\ref{Eqs:ModBloch}) together with Eq.~(\ref{Eq:Mxchi}) yields
\bse
\label{Eqs:chippp11}
\bea
\frac{\chi^\prime(\omega)}{\chi_0} &=& \frac{1 + (\omega\tau)^2 +2(\omega_0\tau)^2 - (\omega\tau)^2(\omega_0\tau)^2 + (\omega_0\tau)^4}
{1 + 2 [(\omega\tau)^2 + (\omega_0\tau)^2] + [(\omega\tau)^2 - (\omega_0\tau)^2]^2 },\nonumber\\
\label{Eq:chipchipMod}\\
\frac{\chi^{\prime\prime}(\omega)}{\chi_0} &=& \frac{\omega\tau}{2}\left[\frac{1}{1+(\omega-\omega_0)^2\tau^2} + \frac{1}{1+(\omega+\omega_0)^2\tau^2}\right].\nonumber\\
\label{Eq:chippchipMod}
\eea
\ese
The respective limiting low- and high-frequency behaviors are
\bse
\bea
\frac{\chi^{\prime}(\omega\to0)}{\chi_0} &=& 1 - \frac{1- (\omega_0\tau)^2 }{[1+(\omega_0\tau)^2]^2}(\omega\tau)^2,\\
\frac{\chi^{\prime}(\omega\to\infty)}{\chi_0} &=& \frac{1-(\omega_0\tau)^2}{(\omega\tau)^2},\label{Eq:chipInf}\\
\frac{\chi^{\prime\prime}(\omega\to0)}{\chi_0} &=& \frac{\omega\tau}{1+(\omega_0\tau)^2},\\
\frac{\chi^{\prime\prime}(\omega\to\infty)}{\chi_0} &=& \frac{1}{\omega\tau}.\label{Eq:chippOmToInfty}
\eea
\ese

Due to the unexpected form of $\chi^\prime(\omega)$ in Eq.~(\ref{Eq:chipchipMod}), we checked its validity by deriving $\chi^\prime(\omega)$ using $\chi^{\prime\prime}(\omega)$ in Eq.~(\ref{Eq:chippchipMod}) and the Kramers-Kronig relation~\cite{Slichter1963}
\be
\chi^\prime(\omega)-\chi^\prime(\infty) = \frac{1}{\pi}{\cal P}\int_{-\infty}^\infty \frac{\chi^{\prime\prime}(\omega^\prime)}{\omega^\prime - \omega}d\omega^\prime,
\label{Eq:KK}
\ee
where ${\cal P}$ denotes the principal part of the integral and according to Eq.~(\ref{Eq:chipInf}) $\chi^\prime(\infty) = 0$.  The  resulting $\chi^\prime(\omega)$ was found to be identical with Eq.~(\ref{Eq:chipchipMod}).  Furthermore, the more involved expression for $\chi^\prime(\omega)$ given in Ref.~\cite{Garstens1955} was found to be equivalent to  Eq.~(\ref{Eq:chipchipMod}).

One sees that $\chi^\prime(\omega)$ is even in $\omega$ and $\chi^{\prime\prime}(\omega)$ is odd in $\omega$ as required.  In addition, $\chi^{\prime\prime}$ at $\omega_0=0\ (H_0=0)$ is nonzero when $\omega$ is finite, thus correcting the null value of $\chi^{\prime\prime}(\omega)$ for finite $\omega$ obtained from the Bloch equations for $\omega_0=0$ at the end of Sec.~\ref{Sec:Bloch}.  We note that $\chi^\prime(\omega)$ in Eq.~(\ref{Eq:chipchipMod}) can be rewritten as
\be
\frac{\chi^\prime(\omega)}{\chi_0} = 1 + \frac{\omega\tau}{2}\left[\frac{\omega_0\tau - \omega\tau}{1 + (\omega_0\tau-\omega\tau)^2} - \frac{\omega_0\tau + \omega\tau}{1 + (\omega_0\tau+\omega\tau)^2}\right],
\label{Eq:chipAlt}
\ee
where the two terms in square brackets might be viewed as arising from the rotating- and counter-rotating components of the linearly-polarized ${\bf H}_1$ in Eqs.~(\ref{Eq:RC}) and~(\ref{Eq:CRC}).  However, this identification is misleading because the additive factor of unity on the right side of Eq.~(\ref{Eq:chipAlt}) is also part of $\chi^\prime(\omega)$.

The half width at half maximum peak height~$\Delta\omega$ (HWHM) of $\chi^{\prime\prime}(\omega)$ in Eq.~(\ref{Eq:chippchipMod}) is given in terms of $\tau$ by Eq.~(\ref{Eq:T2Dw}) with $\tau$ replacing~$T_2$.  With this identification, Eqs.~(\ref{Eqs:chippp11}) become
\bse
\label{Eqs:chipchippMod2}
\bea
\frac{\chi^\prime(\omega)}{\chi_0} &=& \frac{\Delta\omega^4 + \Delta\omega^2(\omega^2 +2 \omega_0^2) -  \omega^2\omega_0^2 + \omega_0^4 }
{ \Delta\omega^4 + 2 \Delta\omega^2(\omega^2 + \omega_0^2) + (\omega^2 - \omega_0^2)^2},
\label{Eq:chip24}\\
\frac{\chi^{\prime\prime}(\omega)}{\chi_0} &=& \frac{\omega}{2}\left[\frac{\Delta\omega}{\Delta\omega^2+(\omega-\omega_0)^2} + \frac{\Delta\omega}{\Delta\omega^2+(\omega+\omega_0)^2}\right].\nonumber\\
\label{Eq:chipp55}
\eea
\ese
Equations~(\ref{Eqs:chipchippMod2}) are appropriate for scanning $\omega$ at fixed dc field $H_0 \equiv \omega_0/\gamma$.

If one scans $H_0$ at fixed resonant frequency $\omega\equiv\omega_{\rm res}$ as is typical in CW EPR experiments, then with the substitutions $\Delta\omega \to \gamma\Delta H$, $\omega \to \gamma H_{\rm res}$, and $\omega_0\to\gamma H$, Eqs.~(\ref{Eqs:chipchippMod2}) become
\bse
\label{Eqs:chi(H)}
\bea
\frac{\chi^\prime(H)}{\chi_0} &=& \label{Eq:chip(H)}\\
&& \hspace{-0.5in}\frac{\Delta H^4 + \Delta H^2(H_{\rm res}^2 + 2H^2) - H_{\rm res}^2H^2 + H^4 }
{\Delta H^4 + 2\Delta H^2(H_{\rm res}^2 + H^2) + (H_{\rm res}^2 - H^2)^2},\nonumber\\
\frac{\chi^{\prime\prime}(H)}{\chi_0} &=& \frac{H_{\rm res}}{2} \label{Eq:chipp(H)}\\
&&\hspace{-0.5in}\times\ \left[\frac{\Delta H}{\Delta H^2+(H_{\rm res}-H)^2} + \frac{\Delta H}{\Delta H^2+(H_{\rm res}+H)^2}\right],\nonumber
\eea
\ese
where the replacement $H_0\to H$ was made in Eq.~(\ref{Eq:H0vec}) so as to conform to the conventional symbol $H$ for the applied dc magnetic field.  The experimental fitting parameters in Eqs.~(\ref{Eqs:chi(H)})  are $\Delta H$ and $H_{\rm res}$.  One sees that both $\chi^\prime(H)$ and $\chi^{\prime\prime}(H)$ are even in~$H$, distinctly different from the case of varying $\omega$ with $H$ held constant where $\chi^{\prime\prime}(\omega)$ is odd in~$\omega$.

The respective low- and high-frequency series expansions of Eqs.~(\ref{Eqs:chi(H)}) are
\bse
\bea
\frac{\chi^{\prime}(H\to0)}{\chi_0} &=& \frac{\Delta H^2}{H_{\rm res}^2+\Delta H^2}
\\
&& -\ \frac{(H_{\rm res}^2-2\Delta H^2)(H_{\rm res}^2-\Delta H^2)}{(H_{\rm res}^2 + \Delta H^2)^3}H^2,\nonumber\\
\frac{\chi^{\prime}(H\to\infty)}{\chi_0} &=& 1 + \frac{H_{\rm res}^2-2\Delta H^2}{H^2},\label{Eq:chipHInf}\\
\frac{\chi^{\prime\prime}(H\to0)}{\chi_0} &=& \frac{H_{\rm res}\Delta H}{H_{\rm res}^2 + \Delta H^2}\\
&& +\ \frac{H_{\rm res}\Delta H(3H_{\rm res}^2 - \Delta H^2)}{(H_{\rm res}^2 + \Delta H^2)^3}H^2\nonumber\\
\frac{\chi^{\prime\prime}(H\to\infty)}{\chi_0} &=& \frac{H_{\rm res}\Delta H}{H^2}.
\eea
\ese
Contrary to the limit $\chi^\prime(\omega\to\infty) = 0$ in Eq.~(\ref{Eq:chipInf}), the corresponding limit of  $\chi^\prime(H\to\infty)$ in Eq.~(\ref{Eq:chipHInf}) is unity.  This nonzero value is not relevant when we compute the field derivative of $\chi_{\rm D}(H)$ that is used to fit the field-derivative of experimental CW EPR spectra as discussed in Sec.~\ref{Eq:DysonEPR} below.

The integrals of $\chi^\prime(H)-\chi^\prime(H = \infty)$ and $\chi^{\prime\prime}(H)$ over nonnegative values of~$H$ are obtained from  Eqs.~(\ref{Eqs:chi(H)}) as
\bse
\label{Eqs:ChipChippIntegrals}
\bea
\int_0^\infty[\chi^\prime(H)-\chi^\prime(\infty)]dH &=& 0,\\
&&\hspace{-1.55in} \int_0^\infty\chi^{\prime\prime}(H)dH = \frac{\pi H_{\rm res}}{2}\chi_0,
\eea
\ese
where $\chi^\prime(\infty)=\chi_0$.

\subsection{\label{Sec:PowerSkinDepth} Power Absorption and Skin Depth}

The time-dependent power absorbed by a resonant system is given by~\cite{Abragam1961,Pake1973}
\be
P(t) = -{\bf M}(t)\cdot \frac{d{\bf H}(t)}{dt}.
\label{Eq:P(t)}
\ee
Using Eq.~(\ref{Eq:LP}) and~(\ref{Eq:Mxchi}) the time-average of $P(t)$ in Eq.~(\ref{Eq:P(t)}) for a volume~$V$ of a sample xexposed to the rf magnetic field  is
\be
P = \frac{H_1^2V}{2}\omega\chi^{\prime\prime}(\omega).
\label{Eq:PvsOmega}
\ee 
The volume~$V$ is not necessarily the volume of the sample if it is conducting, since the skin depth~$\delta$ may limit the volume of the sample exposed to the rf magnetic field.

The expression for $\delta$ is
\be
\delta({\rm m}) = \sqrt{\frac{\rho(\Omega~{\rm m})}{\pi f({\rm Hz}) \mu({\rm H/m})}}\qquad ({\rm SI\ units}),
\label{Eq:delta(m)}
\ee
where $\rho$ is the electrical resistivity, $f$ is the rf frequency, and $\mu$ is the magnetic permeability of the sample. For a sample surface parallel to ${\bf H}_1$ with area~$A$, the volume of the sample exposed to the rf magnetic field is 
\be
V = A\delta
\ee
if the thickness of the sample perpendicular to ${\bf H}_1$ is larger than $\delta$, and the volume of the sample otherwise.  The magnetic permeability in SI units is given by 
\be
\mu = \mu_0[1 + (M/H)],
\label{Eq:mu}
\ee
where $\mu_0$ is the magnetic permeability of free space. $M$ is the volume magnetization of the sample and in general the dimensionless ratio $M(T,H)/H$ can be dependent on the temperature~$T$ and the magnitude and direction of the applied field~${\bf H}$.  The required SI value of $M/H$ is obtained from $M/H$ in dimensionless cgs units via $M/H\ ({\rm SI}) =(4\pi)^{-1}(M/H)$~(cgs).

Using Eqs.~(\ref{Eq:chippchipMod}) and~(\ref{Eq:PvsOmega}), the integrated power absorption $P_{\rm int}(\omega)$ for fixed field~$H$ and varying $\omega$ for frequencies up to $\omega$ is
\begin{widetext}
\bea
P_{\rm int}(\omega) &=& \int_0^\omega P(\omega^\prime)d\omega^\prime \\
&&\hspace{-0.6in} = \frac{H_1^2 V \chi_0}{2}\bigg\{2\omega\Delta\omega + \left(\omega_0^2-\Delta\omega^2\right)\bigg[\arctan\bigg(\frac{\omega-\omega_0}{\Delta\omega}\bigg) + \arctan\bigg(\frac{\omega+\omega_0}{\Delta\omega}\bigg)\bigg]-\ 2\omega_0\Delta\omega\, {\rm  arctanh}\bigg(\frac{2\omega\omega_0}{\omega^2+\omega_0^2+\Delta\omega^2}\bigg)\bigg\}.\nonumber
\eea
\end{widetext}
The limiting behaviors are
\bse
\bea
P_{\rm int}(\omega\to0) &=& H_1^2 V \chi_0\left[\frac{\omega^3\Delta\omega}{3(\omega_0^2 + \Delta\omega^2)}\right],\label{Eq:PConstH}\\
P_{\rm int}(\omega\to\infty) &=&   H_1^2 V \chi_0\left[\omega\,\Delta\omega + \frac{\pi}{2}(\omega_0^2-\Delta\omega^2)\right].\nonumber\\
\label{Eq:PintConstH}
\eea
\ese
Thus $P_{\rm int}(\omega)$ is proportional to $\omega^3$ at low frequencies and diverges linearly with $\omega$ for \mbox{$\omega\gg \omega_0^2/\Delta\omega,\,\Delta\omega$}.

When $H$ is scanned at constant $\omega=\omega_{\rm res} = \gamma H_{\rm res}$, one obtains
\be
P(H) =H_1^2V\gamma H_{\rm res}\chi^{\prime\prime}(H),
\ee
where $\chi^{\prime\prime}(H)$ is given in Eq.~(\ref{Eq:chipp(H)}).  The integrated power versus~$H$ is now given by
\bea
P_{\rm int}(H) &=& \int_0^H P(H^\prime)dH^\prime \\
&=& \frac{\chi_0 H_1^2 H_{\rm res}^2V\gamma}{2}\bigg[\arctan\left(\frac{H-H_{\rm res}}{\Delta H}\right)\nonumber\\
&& +\ \arctan\left(\frac{H+H_{\rm res}}{\Delta H}\right)\bigg].\nonumber
\eea
The limiting behaviors of $P_{\rm int}(H)$ are
\bse
\bea
P_{\rm int}(H\to0) &=& \chi_0 H_1^2 H_{\rm res}^2V\gamma\left(\frac{H\Delta H}{H_{\rm res}^2 + \Delta H^2}\right),\label{Eq:PintHto0}\hspace{0.4in}\\
P_{\rm int}(H\to\infty) &=& \chi_0 H_1^2 H_{\rm res}^2V\gamma\left(\frac{\pi}{2} - \frac{\Delta H}{H}\right).\label{Eq:PintHtoinfty}
\eea
\ese
Thus for fixed~$\omega$ and varying~$H$, at low fields the absorbed power is proportional to $H$ and the high-field limit of $P_{\rm int}$ in Eq.~(\ref{Eq:PintHtoinfty}) is finite in contrast to the diverging behavior in Eq.~(\ref{Eq:PintConstH}) for the high-frequency limit of $P_{\rm int}(\omega)$ with fixed~$H$\@.

\subsection{\label{Eq:DysonEPR} Dysonian EPR of Local Magnetic Moments in Metals}

EPR in metals was studied theoretically by Dyson in 1955 \cite{Dyson1955} and his predictions were first utilized to interpret experimental conduction-electron paramagnetic-resonance data by Feher and Kip \cite{Feher1955}.  A different case of Dyson's theory describes EPR of well-defined local magnetic moments in metals, where the Dysonian absorptive susceptibility $\chi_{\rm D}^{\prime\prime}(\omega)$ contains a contribution from the dispersive susceptibility $\chi^\prime(\omega)$ according to Eqs.~(\ref{Eq:chiDpp}), where $0\leq\alpha\leq1$ with $\alpha=0$ or~1 if the rf skin depth $\delta$ is much larger than or much smaller than the sample dimension that is perpendicular to the linearly-polarized ${\bf H}_1$ in Eq.~(\ref{Eq:LP}), respectively.  Since we obtain EPR spectra versus $H$ at constant~$\omega$, one has
\be
\chi_{\rm D}^{\prime\prime}(H) = \chi^{\prime\prime}(H) + \alpha \chi^{\prime}(H),
\label{Eq:chiDpp10}
\ee
where $\chi^\prime(H)$ and  $\chi^{\prime\prime}(H)$ are given in Eqs.~(\ref{Eqs:chi(H)}).  

EPR of local magnetic moments in metals can only be observed for a limited range of local moments \cite{Taylor1975}, such as for the $s$-state ions Gd$^{+3}$ and Eu$^{+2}$ with $S=7/2$ and $L=0$, where the resonance is observed even in concentrated alloys and compounds.  EPR of the Kramers ions Dy$^{+3}$, Er$^{+3}$, and Yb$^{+3}$ in alloys and compounds have also been observed. Among the $3d$ transition elements, EPR spectra of Mn$^{+2}$ with electron configuration $3d^5$ with $S=5/2$ have also been obtained.  Generally fine and hyperfine features are not resolved in the EPR spectra for high concentrations of these ions in metallic alloys and compounds, where broad featureless Lorentzian-like resonances are observed instead.

\subsection{\label{Sec:PreviousLinshapes} Previously-Used Expressions of $\chi(H)$ for Fitting to Dysonian EPR Spectra}

\begin{figure}
\includegraphics [width=3.3in]{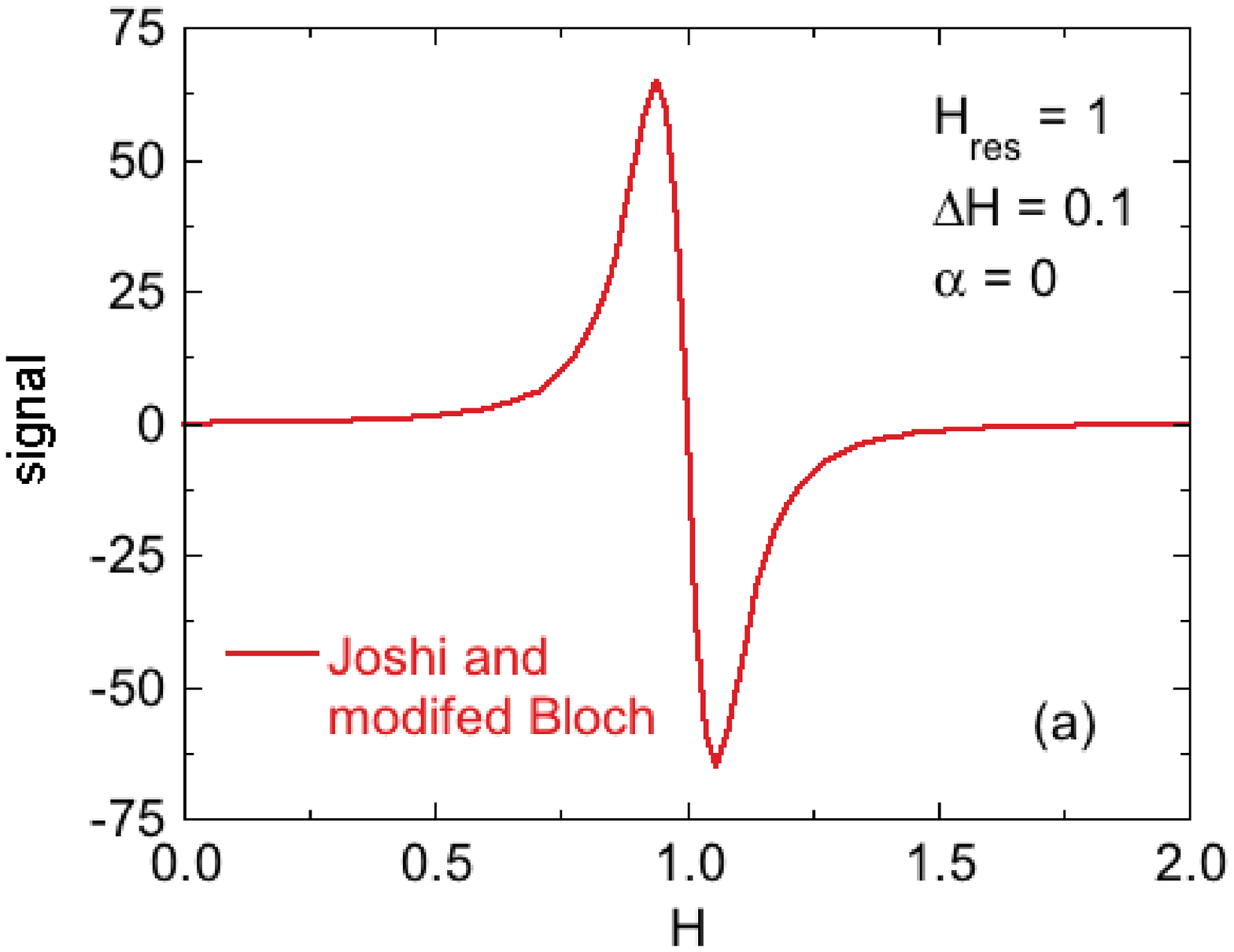}
\includegraphics [width=3.3in]{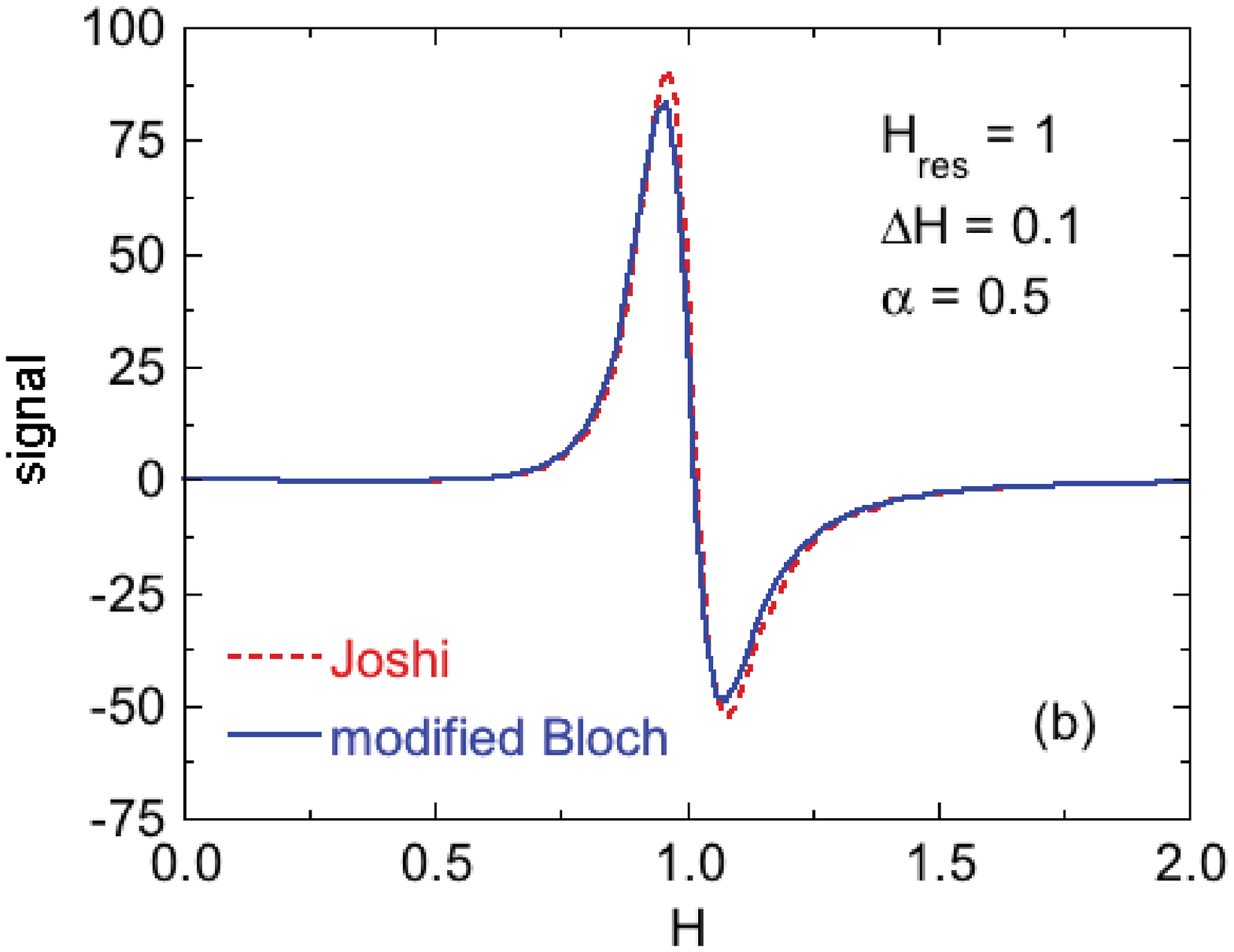}
\includegraphics [width=3.3in]{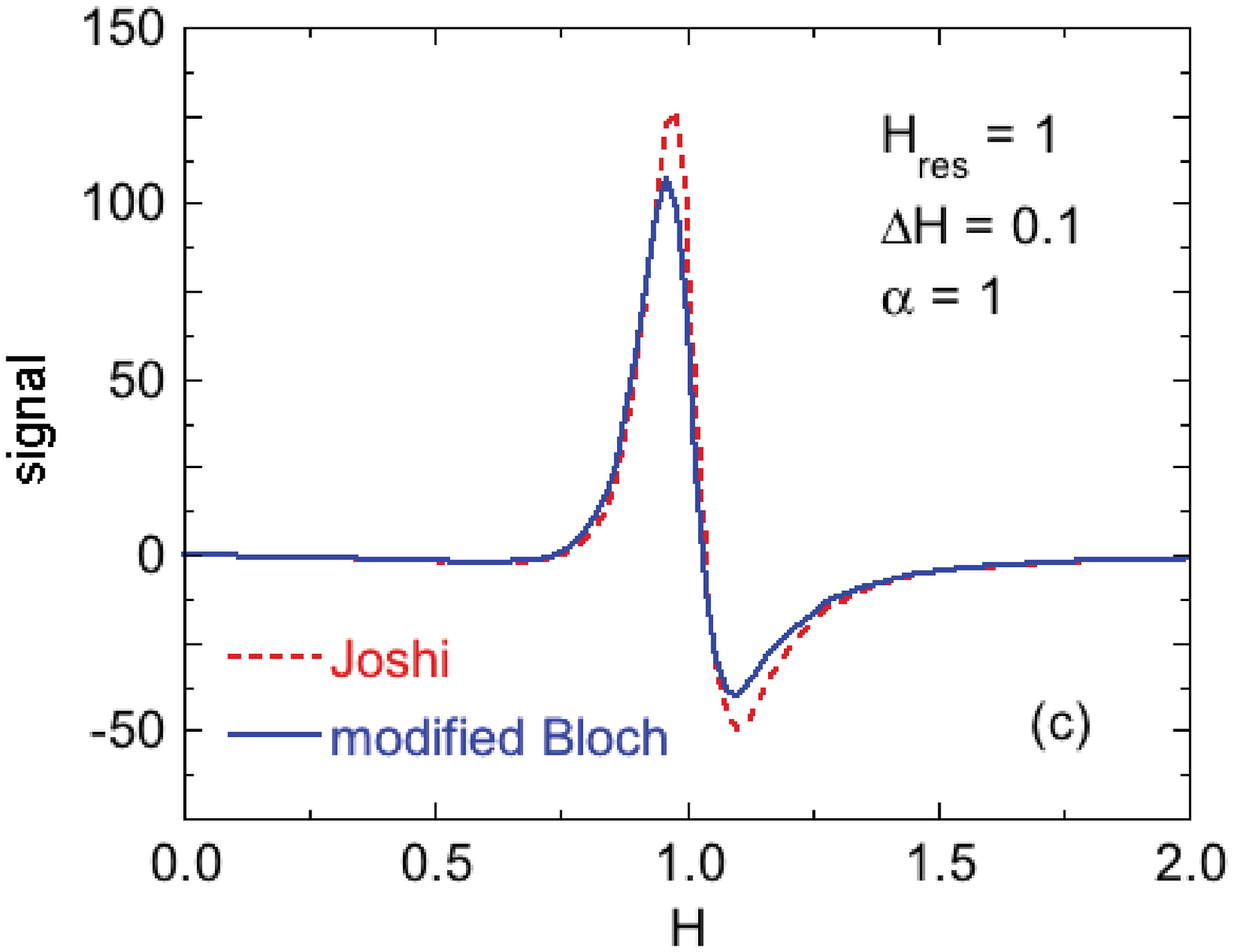}
\caption {Dysonian lineshapes with resonant field $H_{\rm res} = 1$, Lorentzian half width $\Delta H=0.1$, and for Dysonian parameters (a)~$\alpha = 0$, (b)~$\alpha = 0.5$, and (c)~$\alpha = 1$.  The Dysonian lineshapes in Eq.~(\ref{Eq:chiDpp}) using  Eqs.~(\ref{Eqs:ChipppJoshi}) proposed by Joshi and Bhat \cite{Joshi2004} are shown as dashed red curves, whereas our lineshapes using Eqs.~(\ref{Eqs:chi(H)}) are shown as solid blue curves.}
\label{Fig:DysonSpectVSH_DeltH_alpha1}
\end{figure}

\begin{figure}
\includegraphics [width=3.3in]{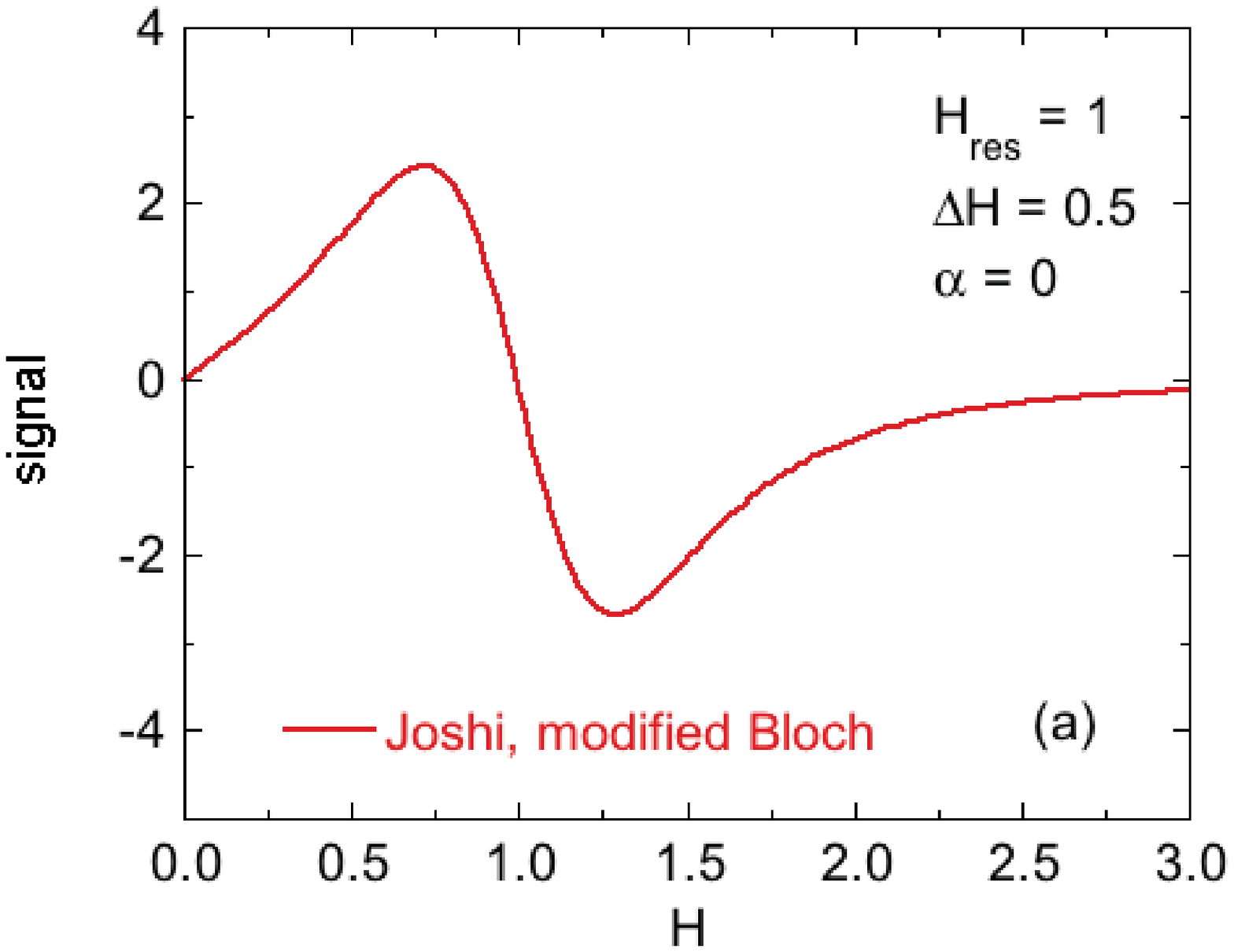}
\includegraphics [width=3.3in]{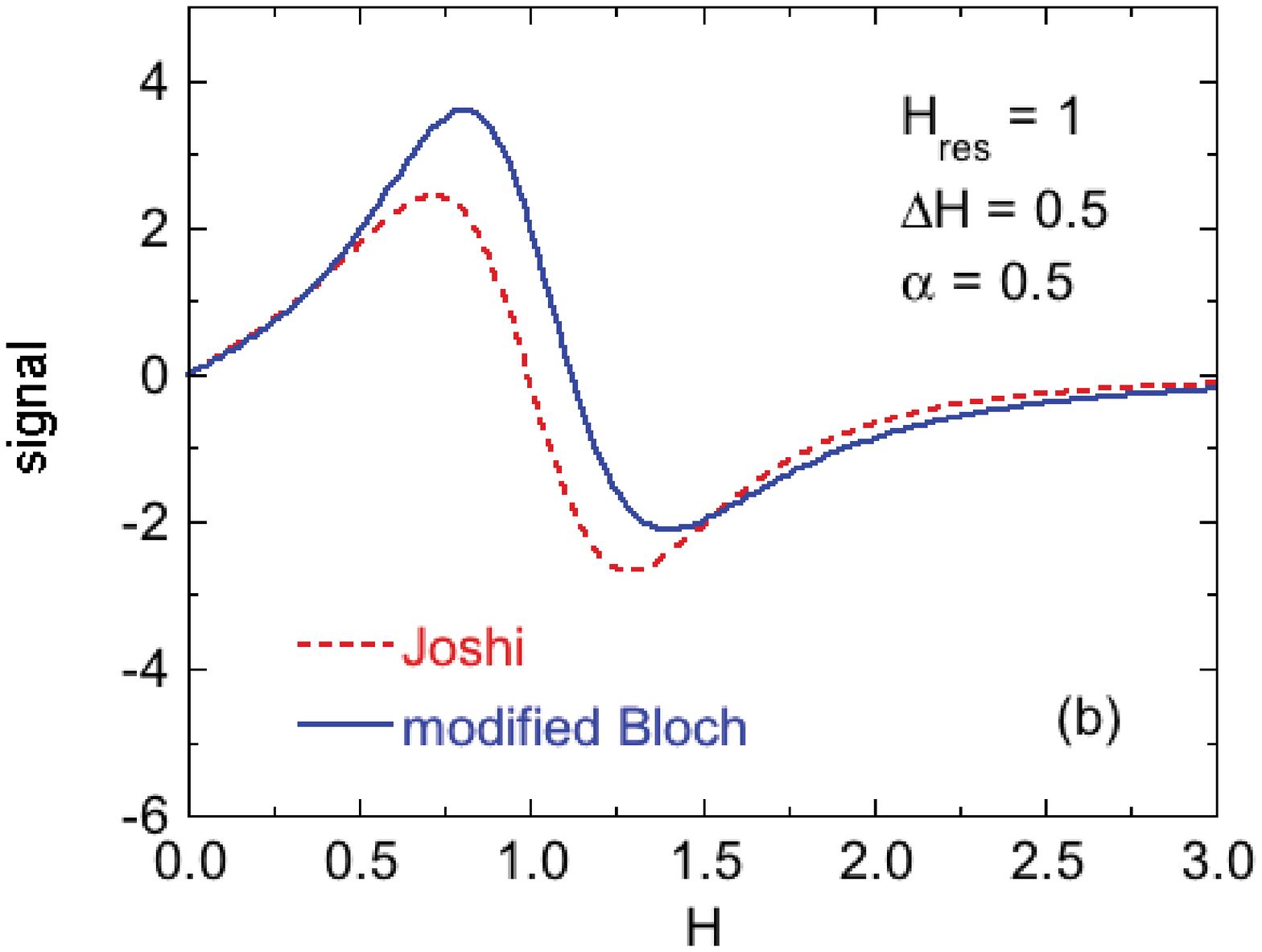}
\includegraphics [width=3.3in]{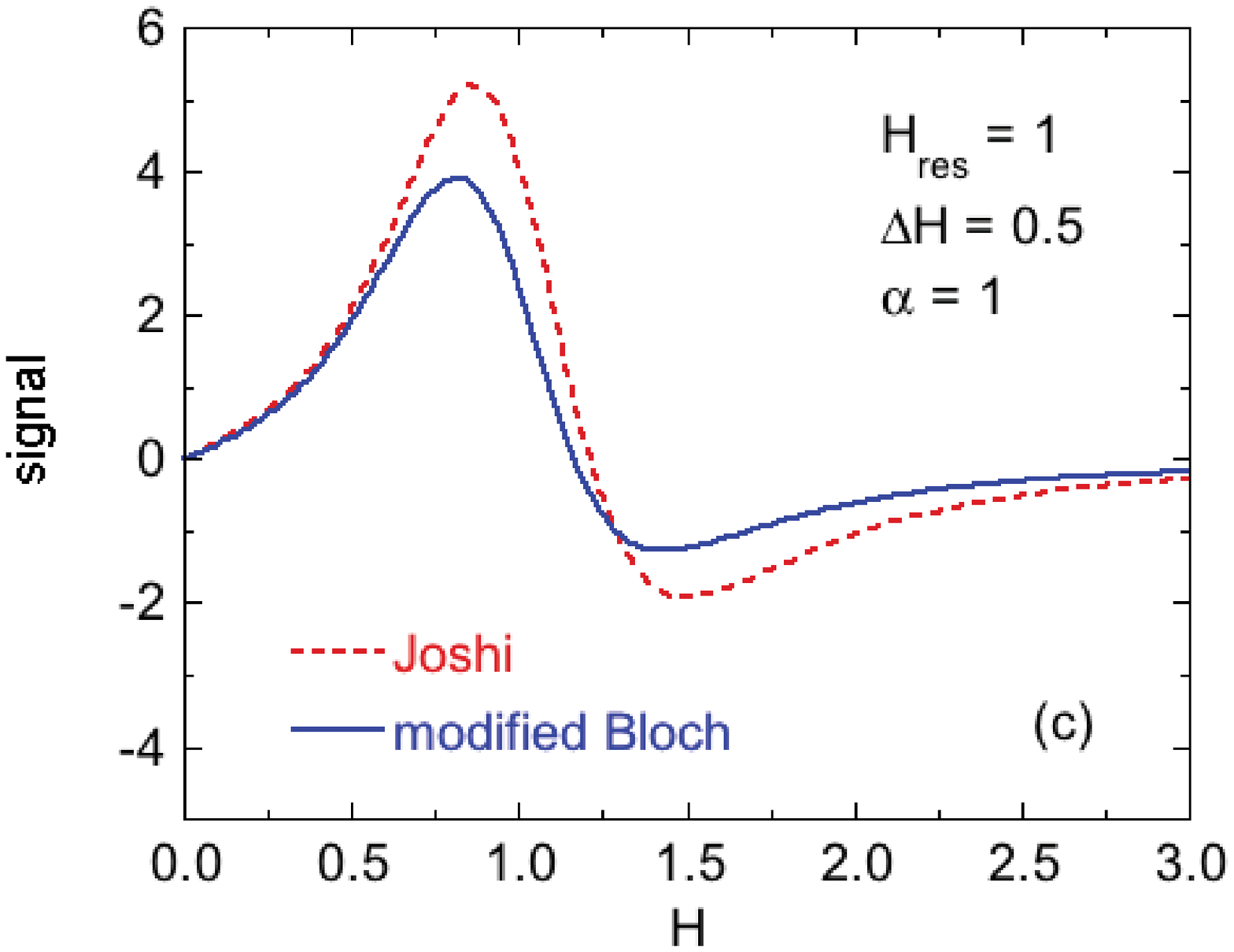}
\caption{Same as Fig.~\ref{Fig:DysonSpectVSH_DeltH_alpha1} but with $\Delta H=0.5$.}
\label{Fig:DysonSpectVSH_DeltH_alpha2}
\end{figure}

The expressions for $\chi^\prime(H)$ and~$\chi^{\prime\prime}(H)$ proposed in Ref.~\cite{Ivanshin2000} to fit Dysonian EPR spectra obtained by sweeping $H$ at fixed~$\omega$ are
\bea
\chi^\prime(H) &\propto& \frac{H-H_{\rm res}}{\Delta H^2+(H_{\rm res}-H)^2} + \frac{H+H_{\rm res}}{\Delta H^2+(H_{\rm res}+H)^2},\nonumber\\
\\
\chi^{\prime\prime}(H) &\propto& \frac{\Delta H}{\Delta H^2+(H_{\rm res}-H)^2} + \frac{\Delta H}{\Delta H^2+(H_{\rm res}+H)^2}.\nonumber
\eea
The function $\chi^\prime(H)$ is odd in $H$ which is not correct, whereas $\chi^{\prime\prime}(H)$ is even in~$H$\@.  The authors of Ref.~\cite{Joshi2004} proposed the modification
\bse
\label{Eqs:ChipppJoshi}
\be
\chi^\prime(H) \propto \frac{H-H_{\rm res}}{\Delta H^2+(H_{\rm res}-H)^2} - \frac{H+H_{\rm res}}{\Delta H^2+(H_{\rm res}+H)^2},\label{Eq:chipJoshi}
\ee
\be
\chi^{\prime\prime}(H) \propto \frac{\Delta H}{\Delta H^2+(H_{\rm res}-H)^2} + \frac{\Delta H}{\Delta H^2+(H_{\rm res}+H)^2}, \label{Eq:chippJoshi}
\ee
\ese
where now both $\chi^\prime(H)$ and $\chi^{\prime\prime}(H)$ are even functions of $H$ as found above in Eqs.~(\ref{Eqs:chi(H)}).

\begin{figure}
\includegraphics [width=3.3in]{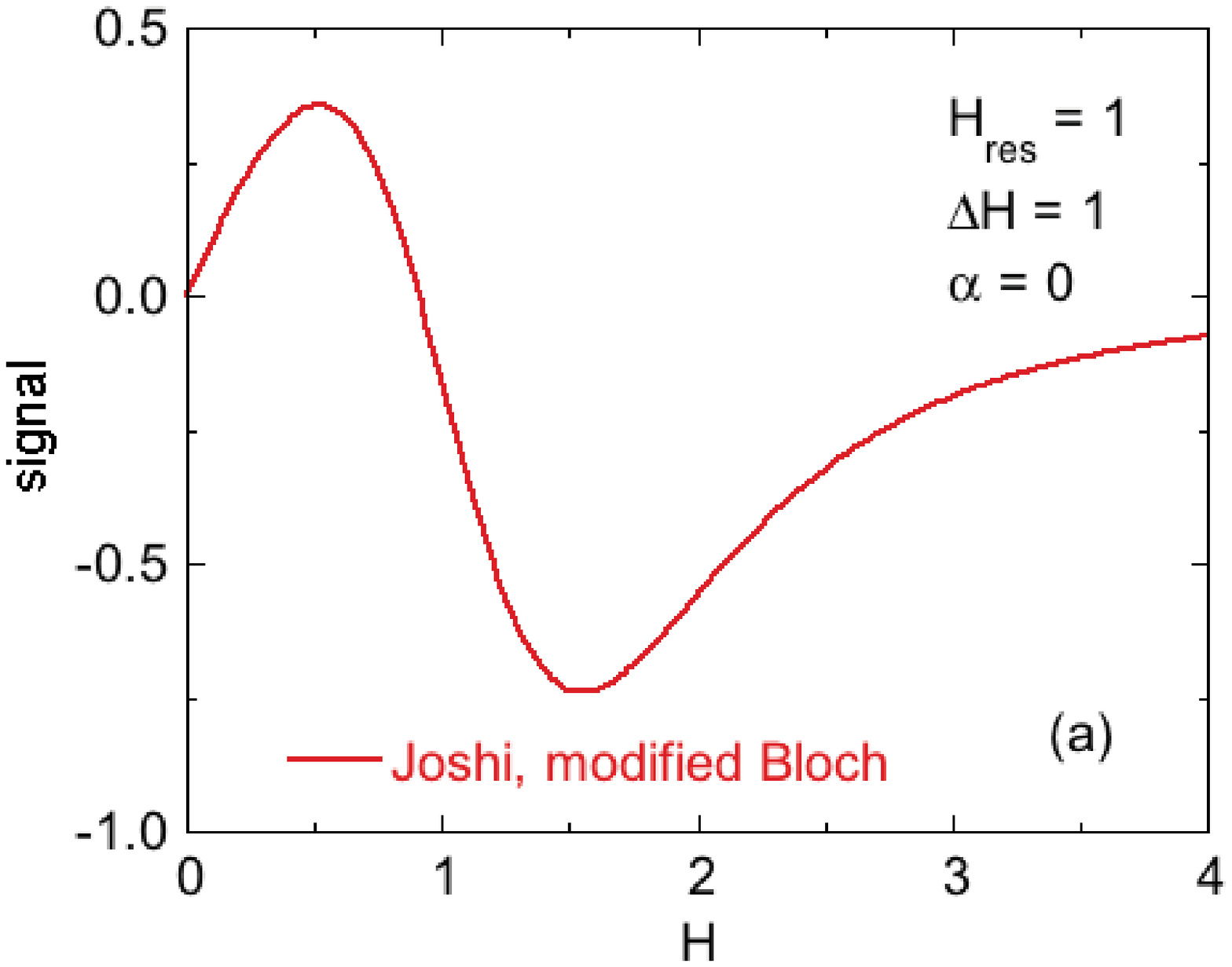}
\includegraphics [width=3.3in]{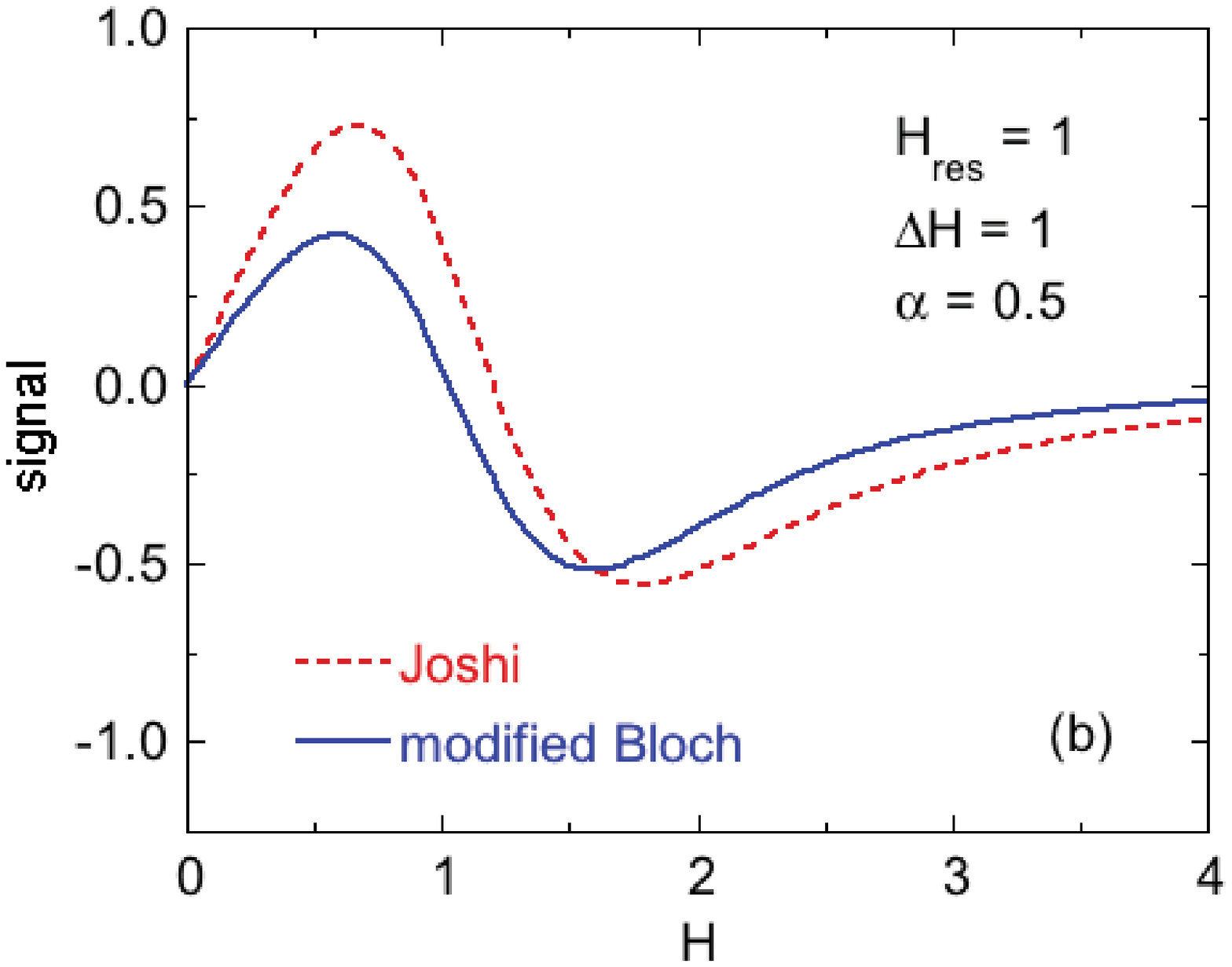}
\includegraphics [width=3.3in]{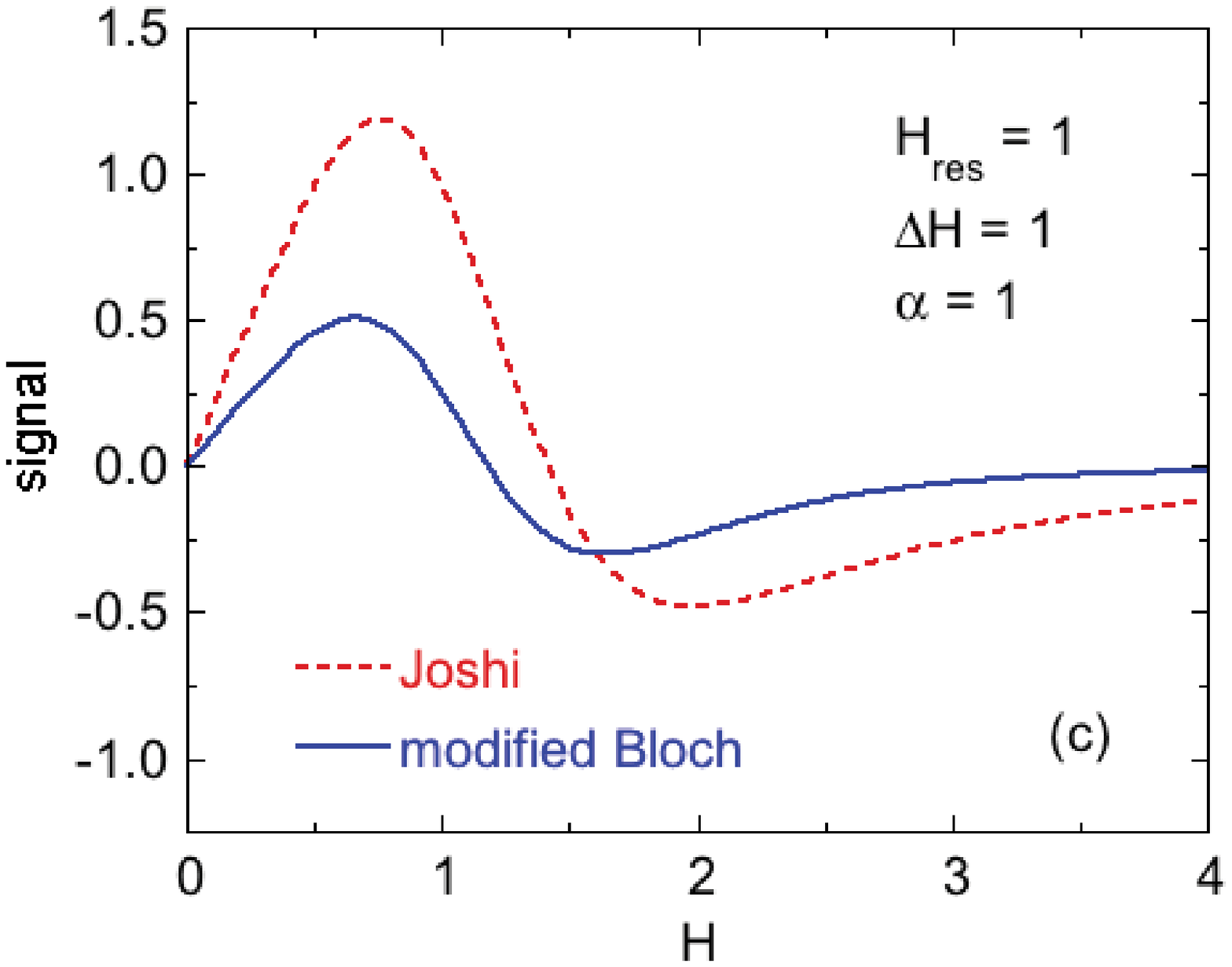}
\caption{Same as Fig.~\ref{Fig:DysonSpectVSH_DeltH_alpha1} but with $\Delta H=1$.}
\label{Fig:DysonSpectVSH_DeltH_alpha3}
\end{figure}

The expression for $\chi^{\prime\prime}(H)$ in Eq.~(\ref{Eq:chippJoshi}) is the same as our result in Eq.~(\ref{Eq:chipp(H)}).  However, the expression for $\chi^\prime(H)$ in Eq.~(\ref{Eq:chipJoshi}) is rather different from that in Eq.~(\ref{Eq:chip(H)}).  Therefore one may expect differences in the fitted field-derivative EPR lineshape parameters when using Eq.~(\ref{Eq:chip(H)}) instead of Eq.~(\ref{Eq:chipJoshi}) for $\chi^\prime(H)$ in the Dysonian $\chi^{\prime\prime}_{\rm D}(H)$  in Eq.~(\ref{Eq:chiDpp10}) with $\alpha>0$. This is confirmed in plots of $\chi_{\rm D}^{\prime\prime}(H)$ in Eq.~(\ref{Eq:chiDpp10}) for $H_{\rm res} = 1$~unit with $\alpha=0$, 0.5, and 1 in Fig.~\ref{Fig:DysonSpectVSH_DeltH_alpha1} with $\Delta H/H_{\rm res} = 0.1$, and similarly in Fig.~\ref{Fig:DysonSpectVSH_DeltH_alpha2} with $\Delta H/H_{\rm res} = 0.5$, and in Fig.~\ref{Fig:DysonSpectVSH_DeltH_alpha3} with $\Delta H/H_{\rm res} = 1$. As $\alpha$ increases from 0 to 0.5 to 1 and as $\Delta H/H_{\rm res}$ increases from 0.1 to~1, one indeed sees a growing divergence between our predictions in Eqs.~(\ref{Eqs:chi(H)}) and those of Joshi and Bhat \cite{Joshi2004} in Eqs.~(\ref{Eqs:ChipppJoshi}).

\begin{figure}
\includegraphics [width=3.3in]{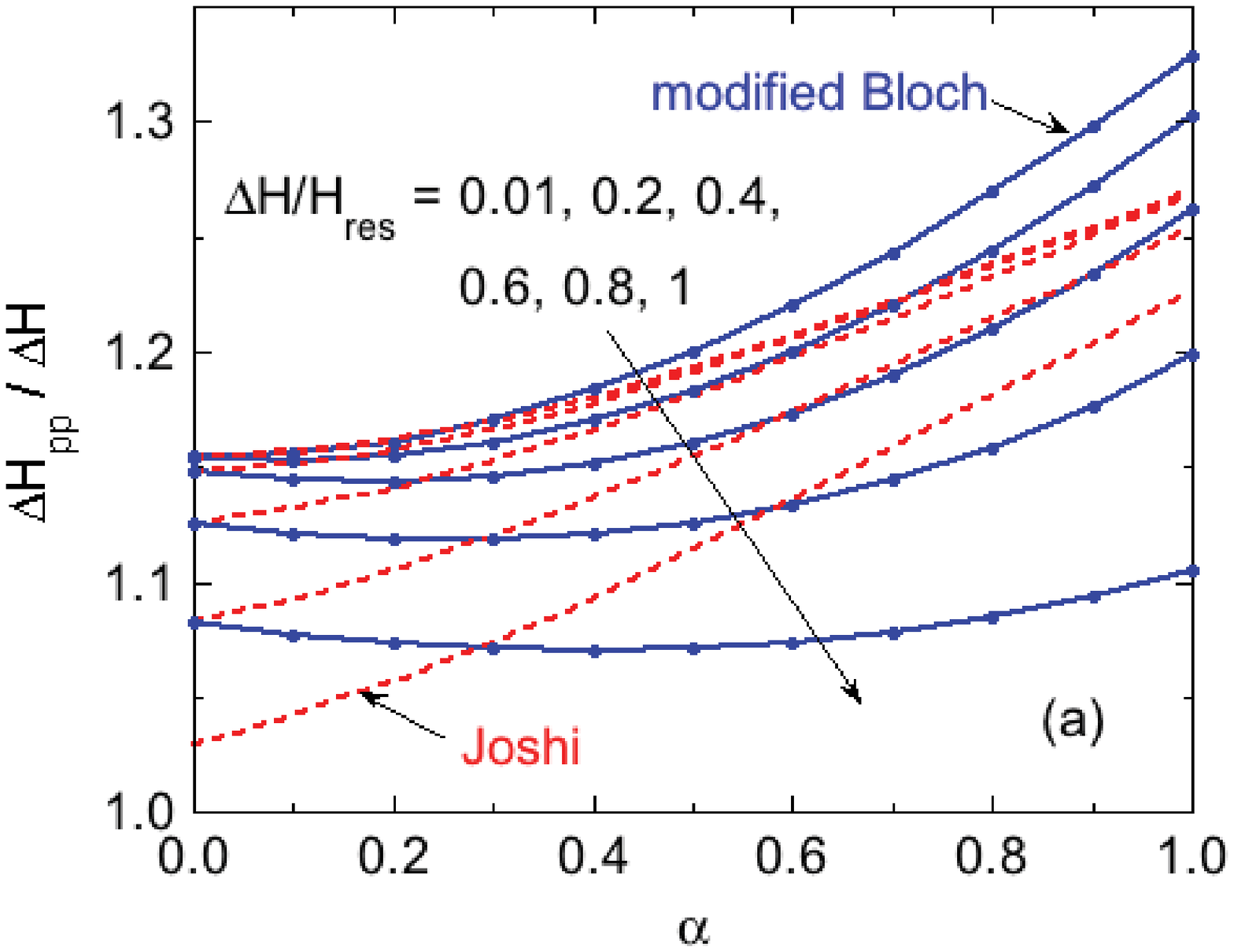}
\includegraphics [width=3.3in]{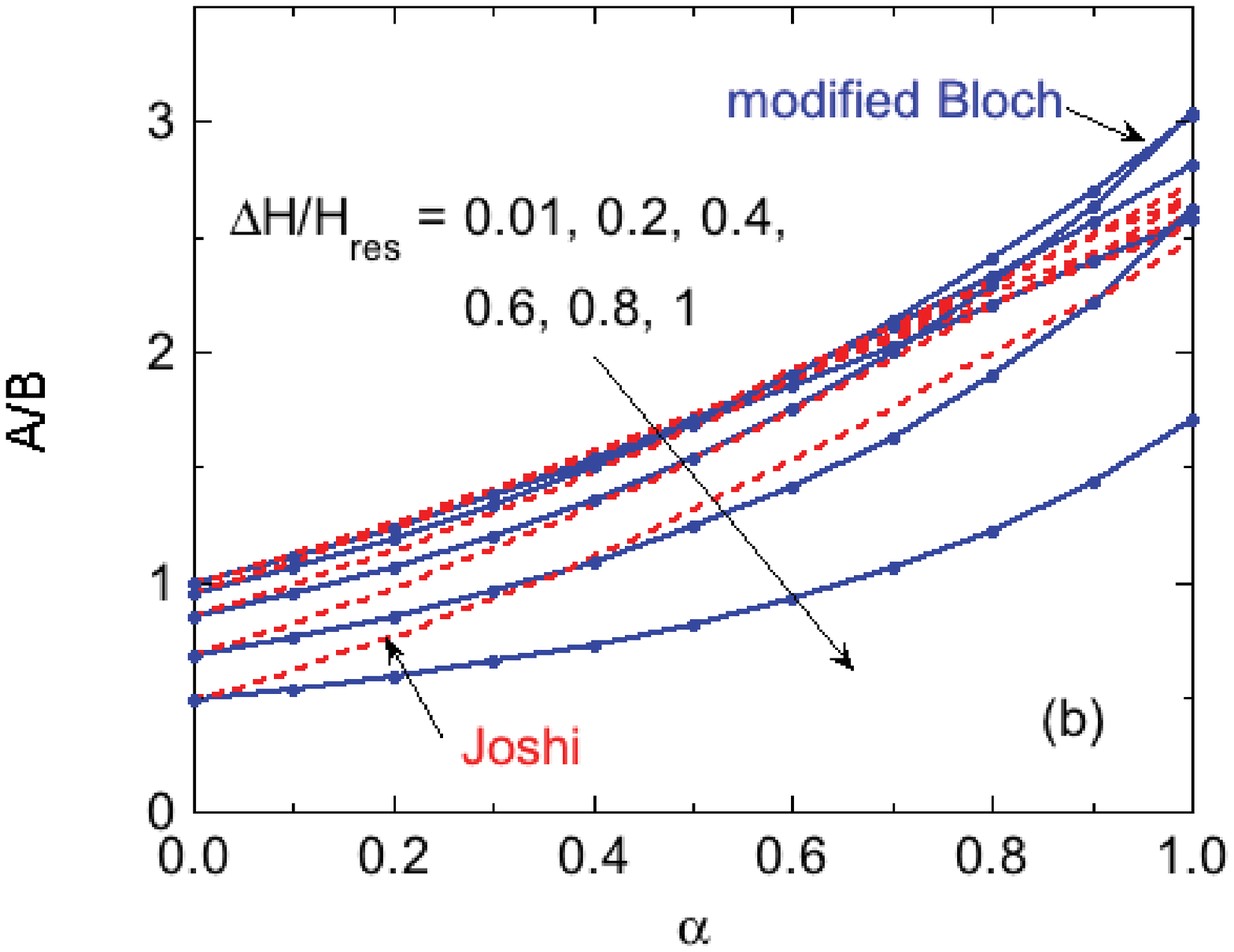}
\caption {Comparison of field-derivative lineshape parameters predicted by Eq.~(\ref{Eq:chiDpp10}) versus the Dysonian lineshape parameter~$\alpha$ using Eqs.~(\ref{Eqs:ChipppJoshi}) proposed by Joshi and Bhat \cite{Joshi2004} (dashed red curves) and from our lineshapes using Eqs.~(\ref{Eqs:chi(H)}) (solid blue curves) for the ratios $\Delta H/H_{\rm res}$ of the Lorentzian half-width $\Delta H$ to the resonant field $H_{\rm res}$ listed in the figures.    (a)~The peak-to-peak linewidth $\Delta H_{\rm pp}$ in the field-derivative spectra divided by  $\Delta H$ versus~$\alpha$ and (b)~the ratio $A/B$ of the first to the second peak heights in the field-derivative spectra versus~$\alpha$.}
\label{Fig:Field-derivative_lineshapes}
\end{figure}

To quantify this divergence, shown in Fig.~\ref{Fig:Field-derivative_lineshapes} are plots versus $\alpha$ of the ratio of the peak-to-peak field width $\Delta H_{\rm pp}$ of derivative spectra such as in Figs.~\ref{Fig:DysonSpectVSH_DeltH_alpha1} to~\ref{Fig:DysonSpectVSH_DeltH_alpha3}  normalized by the Lorentian half width~$\Delta H$ [Fig.~\ref{Fig:Field-derivative_lineshapes}(a)] and of the corresponding $A/B$ ratios of the first to the second peak heights [Fig.~\ref{Fig:Field-derivative_lineshapes}(b)] obtained using our Eqs.~(\ref{Eqs:chi(H)}) as compared with the data obtained using~Eqs.~(\ref{Eqs:ChipppJoshi}). Rather large differences are seen both with increasing $\alpha$ and increasing $\Delta H/H_{\rm res}$.

\begin{figure}[h]
\includegraphics [width=3.3in]{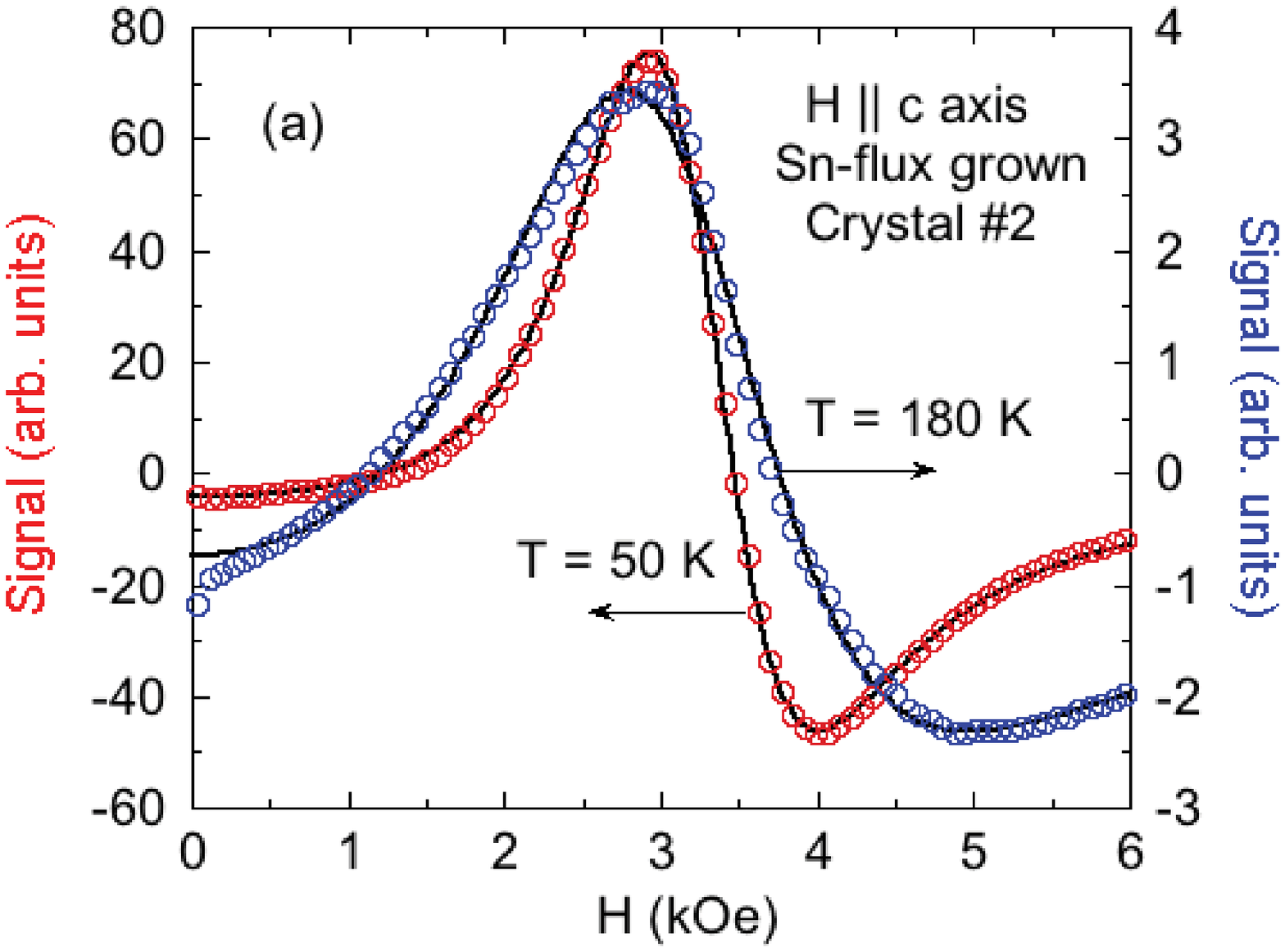}
\includegraphics [width=3.3in]{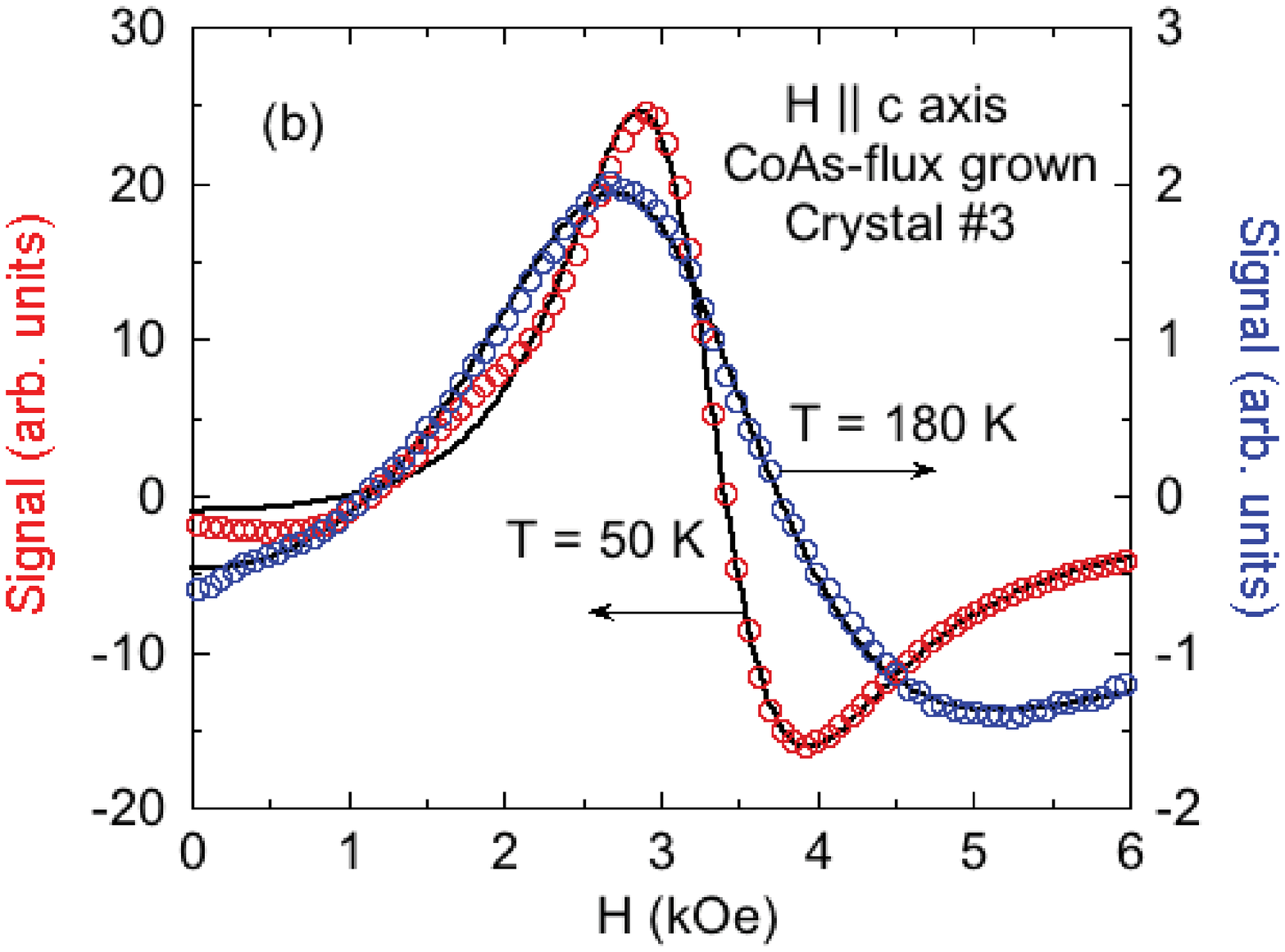}
\caption {Eu$^{+2}$ field-derivative EPR spectra (open circles) at temperatures of 50~K and 180~K for (a)~a Sn-flux-grown crystal~\#2 and (b)~a CoAs self-flux-grown crystal~\#3 of \eca.  The fits by Eq.~(\ref{Eq:dsignaldH}) are shown as solid black curves.}
\label{Fig:SNS_Spectra_Fit}
\end{figure}

\clearpage

\section{\label{Sec:Results} Experimental Results and Analyses}

\subsection{\label{Sec:EPR Overview} EPR Spectra Overview}

Our CW EPR spectrometer is operated under conditions such that the measured unsaturated signal is proportional to $\chi^{\prime\prime}_{\rm D}(H)$.  Experimentally, field modulation and lock-in amplifier detection are used to increase the signal-to-noise ratio in the spectra.  This  means that the field derivative of the Dysonian absorption spectra is measured.  Therefore we fit our spectra by 
\be
\frac{d\,{\rm signal}}{dH} = a + b \frac{d[\chi_{\rm D}^{\prime\prime}(H)/\chi_0]}{dH},
\label{Eq:dsignaldH}
\ee
where $a$ is an instrumental zero offset, $b$ is the amplitude of the signal, and $\chi_{\rm D}^{\prime\prime}(H)/\chi_0$ is given by Eqs.~(\ref{Eqs:chi(H)}) and~(\ref{Eq:chiDpp10}).  The other fitting parameters at each temperature are $\Delta H$, $\alpha$, and~$H_{\rm res}$.

The field-derivative spectra for the two crystals~\#2 and~\#3 of \eca, each at $T=50$~K and 180~K, are shown in Fig.~\ref{Fig:SNS_Spectra_Fit} and exhibit Dysonian lineshapes.  The respective fits by Eq.~(\ref{Eq:dsignaldH}) are shown as solid black curves.  For both crystals, the fits are very good, except for an additional feature for Crystal~\#3 in Fig.~\ref{Fig:SNS_Spectra_Fit}(b) at \mbox{$H \sim 1$~kOe} that disappears above $\sim70$~K\@. As shown in Fig.~\ref{Fig:EuCo2As2c_Hres}(a) below, when this signal disappears the resonance field exhibits a clear discontinuity in the slope versus temperature at $\approx 65$~K\@.  Thus we infer that the signal at $\sim1$~kOe that appears below 65~K in Fig.~\ref{Fig:SNS_Spectra_Fit}(b) for Crystal~\#3 likely does not arise from PM impurities, but is rather associated with some type of second-order phase transition at $\approx 65$~K in this crystal.

From Fig.~\ref{Fig:SNS_Spectra_Fit}, one sees that the linewidth increases with increasing~$T$ for each of the two crystals.  Indeed, the resonances for the two crystals at $T=180$~K are cut off due to the $H=6$~kOe upper limit of our measurements.  From the different ordinate scales for the two temperatures in each of Figs.~\ref{Fig:SNS_Spectra_Fit}(a) and~\ref{Fig:SNS_Spectra_Fit}(b), the signal amplitude strongly decreases with increasing temperature for each crystal.  Such a decrease is expected from the Curie-Weiss $T$ dependence~\cite{Sangeetha2017} of the Eu$^{+2}$ spin susceptibility $\chi_0$ in Eqs.~(\ref{Eqs:chi(H)}).

Using Eqs.~(\ref{Eqs:ChipChippIntegrals}), the double integral of the second term in Eq.~(\ref{Eq:dsignaldH}) over all nonnegative fields~$H$ is
\be
\int_0^\infty dH \int_0^H b\,\frac{d[\chi_{\rm D}^{\prime\prime}(H^\prime)]}{dH^\prime} dH^\prime = \frac{b\pi H_{\rm res}}{2}\chi_0.
\ee
This double integral is thus proportional to the static magnetic susceptibility $\chi_0$ that would be measured using a dc magnetometer, a result that is  often utilized in the literature when discussing the results of EPR measurements of local magnetic moments in metals.

\subsection{\label{Sec:alpha} Dysonian $\alpha$ Parameter}

\begin{figure}
\includegraphics [width=3.3in]{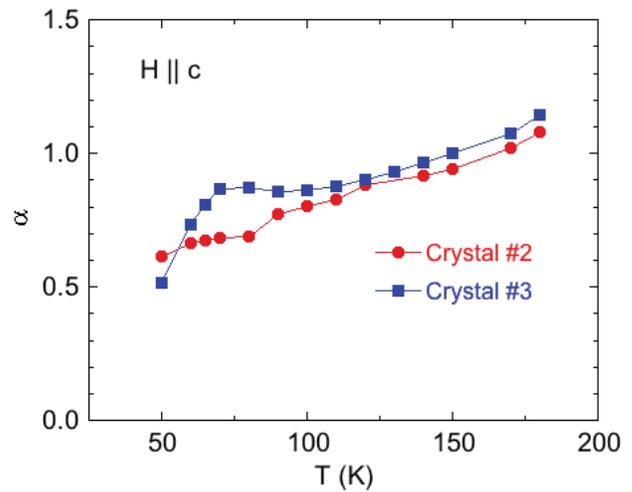}
\caption {Ratio~$\alpha$ of the dispersive susceptibility $\chi^\prime$ to the absorptive susceptibility~$\chi^{\prime\prime}$ in the Dysonian absorptive susceptibility $\chi_{\rm D}^{\prime\prime}(H)$ in Eq.~(\ref{Eq:dsignaldH}) for Crystals~\#2 and \#3 versus temperature~$T$.}
\label{Fig:EuCo2As2c_alpha}
\end{figure}

Shown in Fig.~\ref{Fig:EuCo2As2c_alpha} are plots of $\alpha$ versus~$T$ obtained from fits to the field derivative of the EPR spectra such as in Fig.~\ref{Fig:SNS_Spectra_Fit}.  The value of~$\alpha$ is expected to be in the range $0\leq\alpha\leq1$ for a physically-valid fit as noted previously.  This criterion is satisfied except for the data at the highest temperatures of 170~K and 180~K, which are slightly larger than unity.

For \eca\ in the field and PM temperature ranges of interest in this paper, $M/H$ in Eq.~(\ref{Eq:mu}) is just the magnetic susceptibility per unit volume~$\chi_{\rm V}$ that is given by the Curie-Weiss law in cgs units as
\bse
\label{Eqs:chiV}
\bea
\chi_{\rm V} &=& \frac{C/V_{\rm M}}{T - \theta_{\rm p}},\label{Eq:chiV}\\
C &\approx& 9.0~{\rm \frac{cm^3~K}{mol~Eu}},\\
\theta_{\rm p} &\approx& 22~{\rm K},\\
V_{\rm M} &\approx& 52.6~{\rm \frac{cm^3}{mol~Eu}},
\eea
\ese
where the approximate values of the molar Curie constant~$C$ and the Weiss temperature~$\theta_{\rm p}$ averaged over data for five crystals and over the two field directions ${\bf H}\parallel c$ and \mbox{${\bf H}\parallel ab$} and the molar volume $V_{\rm M}$, all from Ref.~\cite{Sangeetha2017},  are given in the last three of Eqs.~(\ref{Eqs:chiV}).  Then Eq.~(\ref{Eq:chiV}) gives 
\be
\chi_{\rm V}=\frac{0.17~{\rm K}}{T-22~{\rm K}}.
\label{Eq:chiV2}
\ee
At a temperature of 50~K in the PM state where $\chi_{\rm V}$ is near its maximum value versus temperature, the value of $\chi_{\rm V}$ is
\be
\chi_{\rm V}(50~{\rm K}) =0.0060\ ({\rm cgs}),
\ee
so the $M/H$ term in Eq.~(\ref{Eq:mu}) can be set to zero.  Inserting our X-band microwave frequency $f = 9.390$~GHz and the value for $\mu_0$ into Eq.~(\ref{Eq:delta(m)}), one obtains
\be
\delta(T) [\mu{\rm m}] = 0.5194 \sqrt{\rho(T)[\mu\Omega\,{\rm cm}]}.
\label{Eq:delta(m)2}
\ee

When the static field~{\bf H} is applied along the $c$~axis as in this paper, the microwave magnetic field ${\bf H}_1$ is parallel to the $ab$~plane. Since the Poynting vector associated with the skin depth is normal to a surface that is perpendicular to {\bf H}, the microwave electric field associated with ${\bf H}_1$ is also oriented in the $ab$~plane.  Hence the relevant resistivity in Eq.~(\ref{Eq:delta(m)2}) for \mbox{${\bf H}\parallel {\bf c}$-axis} is the in-plane electrical resistivity $\rho_{ab}$, which was measured for two crystals with similar results~\cite{Sangeetha2017} that together are approximated for the temperature range $50~{\rm K}\leq T \leq 300$~K by the linear relation
\be
\rho_{ab}(T) \approx 14~\mu\Omega\,{\rm cm} + \left(0.083\,\frac{\mu\Omega\,{\rm cm}}{\rm K}\right)T. 
\label{Eq:rho(T)}
\ee
\begin{figure}
\includegraphics [width=3.3in]{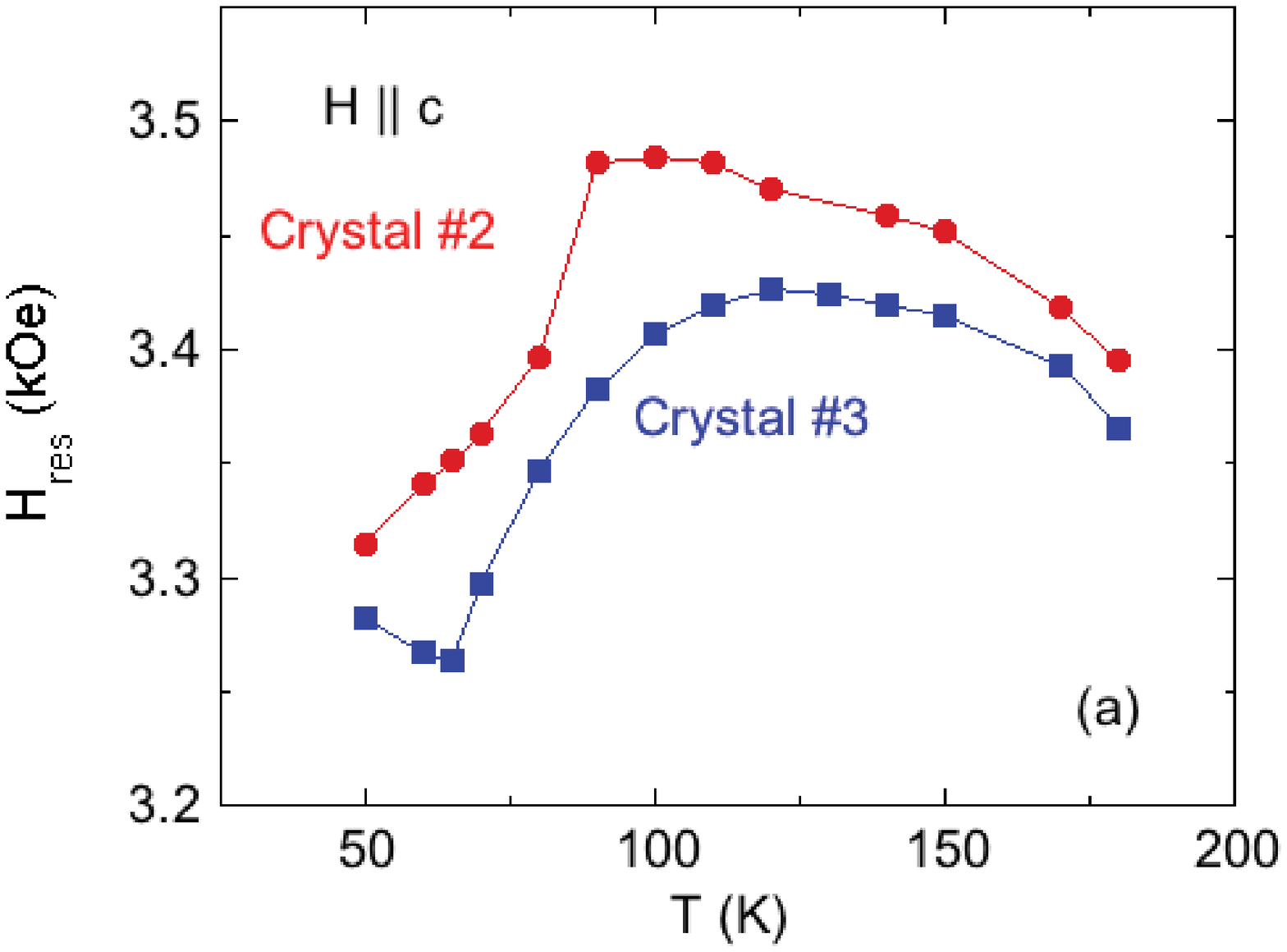}
\includegraphics [width=3.3in]{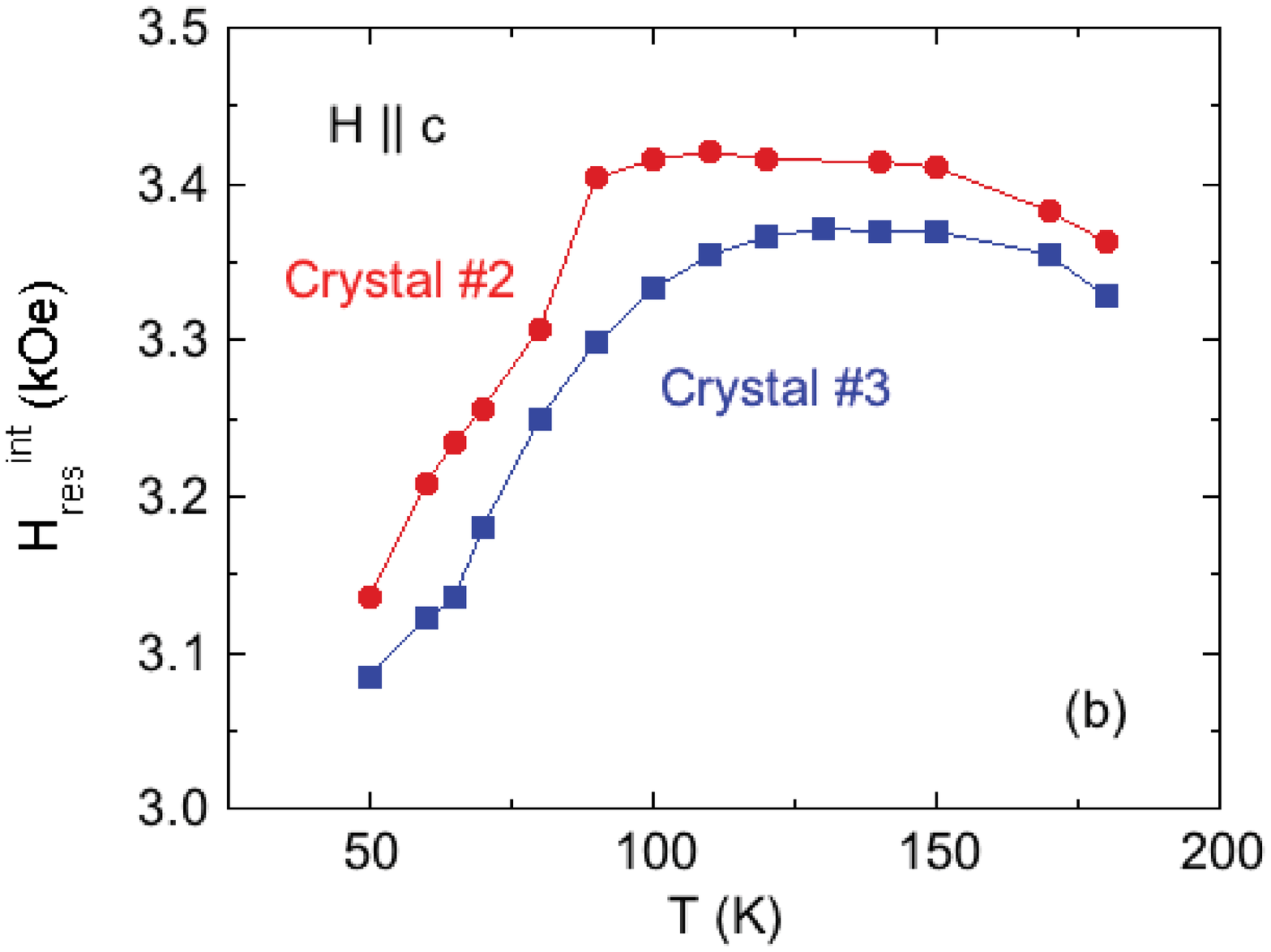}
\includegraphics [width=3.3in]{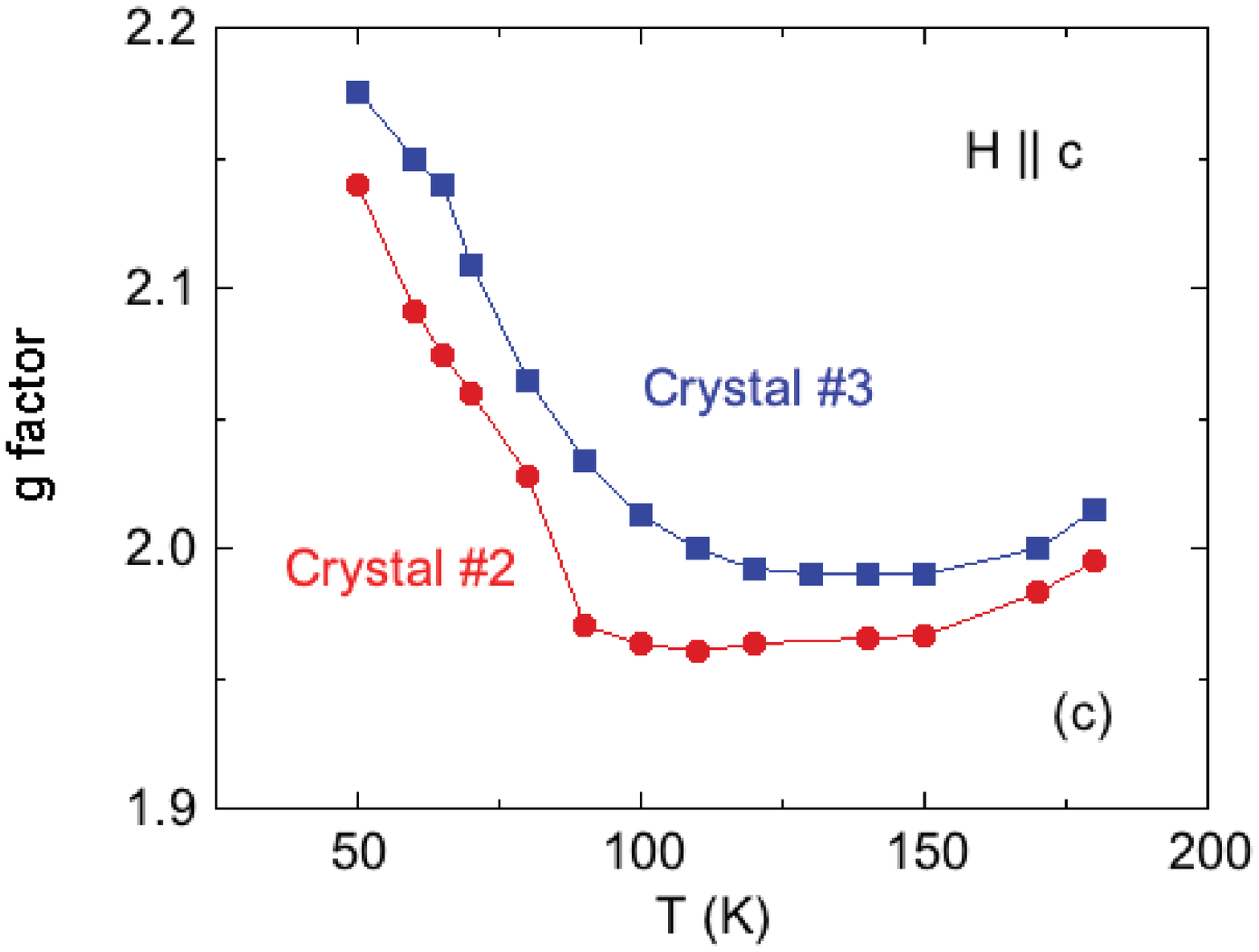}
\caption{(a) The measured Eu$^{+2}$ resonance field $H_{\rm res}$ versus temperature~$T$ for Crystals~\#2 (red circles) and~\#3 (blue squares) of \eca\ obtained from fits of the field-derivative spectra by Eq.~(\ref{Eq:dsignaldH}). (b)~The internal resonance field $H_{\rm res}^{\rm int}$ obtained from the data in panel~(a) after correction for the demagnetizing field according to Eq.~(\ref{Eq:Hintbeta}).  (c)~The Eu$^{+2}$ $g$-factors for Crystals~\#2 and~\#3 versus~$T$ calculated from the data in panel~(b) using Eq.~(\ref{Eq:gCalc}).}
\label{Fig:EuCo2As2c_Hres}
\end{figure}

Using Eqs.~(\ref{Eq:delta(m)2}) and~(\ref{Eq:rho(T)}), one obtains $\delta = 2.2\,\mu$m and $3.2\,\mu$m at 50~K and 300~K, respectively.  These values are much smaller than the $ab$-plane dimensions of our crystals ($\sim$~mm) which would therefore predict $\alpha\approx 1$, as observed at the higher temperatures in Fig.~\ref{Fig:EuCo2As2c_alpha}.    The reason that $\alpha$ decreases with decreasing~$T$ in Fig.~\ref{Fig:EuCo2As2c_alpha} is not clear, especially since $\rho_{ab}$ decreases with decreasing~$T$~\cite{Sangeetha2017}.  This behavior may be related to phase transitions that appear to occur in Fig.~\ref{Fig:EuCo2As2c_alpha} at $T \approx 65$~K for Crystal~\#3 and $\approx 80$~K for Crystal~\#2, as also suggested from the $T$-dependent data for the resonance fields $H_{\rm res}$ for these crystals in Figs.~\ref{Fig:EuCo2As2c_Hres}(a) and~\ref{Fig:EuCo2As2c_Hres}(b) in the following section.

\subsection{\label{Sec:ResFieldg} Resonance Magnetic Field and {\bf g}-Factor}

The resonance magnetic field $H_{\rm res}$ versus temperature obtained from fits of the field-derivative EPR spectra is plotted for Crystals~\#2 and~\#3 in Fig.~\ref{Fig:EuCo2As2c_Hres}(a).  The data show clear evidence for a second-order phase transition in Crystal~\#3 at $\approx 65$~K and either a first- or second-order transition in Crystal~\#2 at $\approx 90$~K\@.

Due to the presence of highly magnetic Eu$^{+2}$ ions with spin $S=7/2$, the magnetic field ${\bf H}^{\rm int}$ internal to a sample can be significantly different from the applied field {\bf H}.  The components of ${\bf H}^{\rm int}$ along the principal-axes directions~$\beta$ are given in Gaussian cgs units by~\cite{Johnston2016}
\bse
\label{Eqs:Hintbeta}
\be
H^{\rm int}_\beta = H_\beta -4\pi N_{{\rm d}\beta} M_\beta,
\label{Eq:Hintbeta}
\ee
where the $\beta$ principal-axis magnetization component $M_\beta$ is in cgs units of Gauss and the magnetometric demagnetization factor $N_{{\rm d}\beta}$ is in SI units where $0\leq N_{{\rm d}\beta}\leq 1$ and \mbox{$\sum_{\beta=1}^3N_{{\rm d}\beta} = 1$}.  The dimensions and $c$-axis $N_{{\rm d}c}$ values of Crystals~\#2 and~\#3 where ${\bf H}\parallel c$~axis were calculated using the expression derived in Ref.~\cite{Aharoni1998} and are listed in Table~\ref{Tab:DemagFac}.

At the field $H_{\rm res}$, Eq.~(\ref{Eq:chiV2}) gives the volume magnetization~$M_c$ as
\be
M_c = \chi_{\rm V}H_{\rm res} = \frac{(0.17~{\rm K})H_{\rm res}}{T-22~{\rm K}}
\label{Eq:M}
\ee
\ese
where 1~G = 1~Oe.  For example, at $T=50$~K, using Eq.~(\ref{Eq:M}) and taking $H_{\rm res} \approx 3.3$~kOe from Fig.~\ref{Fig:EuCo2As2c_Hres}(a) and $N_{{\rm d}c}\approx 0.75$ from Table~\ref{Tab:DemagFac}, one obtains $4\pi N_{{\rm d}c} M_c \approx 0.20$~kOe for this term in Eq.~(\ref{Eq:Hintbeta}), which is similar to the temperature-induced change in $H_{\rm res}$ in Fig.~\ref{Fig:EuCo2As2c_Hres}(a).  Thus taking the demagnetizing field into account results in a significant correction to the measured $H_{\rm res}(T)$ for our crystals.  Shown in Fig.~\ref{Fig:EuCo2As2c_Hres}(b) are data for the internal resonance field $H_{\rm res}^{\rm int}$ versus temperature obtained from the data in Fig.~\ref{Fig:EuCo2As2c_Hres}(a) using Eqs.~(\ref{Eqs:Hintbeta}).  One sees a strong variation of $H_{\rm res}^{\rm int}$ with temperature.

\begin{table}
\caption{\label{Tab:DemagFac} Dimensions and $c$-axis magnetometric demagnetization factors $N_{{\rm d}c}$ of the approximately rectangular-prism-shaped Crystals~\#2 and~\#3.}
\begin{ruledtabular}
\begin{tabular}{cccc}
Crystal								& $ab$-plane 		&  $c$-axis  	& $N_{{\rm d}c}$	\\
               						& (mm$^2$)		&  (mm)						\\
\hline

\#2 EuCo$_{1.99(2)}$As$_2$\footnotemark[1]	& $1.6 \times 6.7$ 	& 0.47	& 0.71				\\
\#3 EuCo$_{1.92(4)}$As$_2$\footnotemark[2]	& $2.5 \times 4.4$ 	& 0.35	& 0.79				\\
\end{tabular}
\end{ruledtabular}
\footnotetext[1]{Grown in Sn flux with H$_2$-treated Co powder}
\footnotetext[2]{Grown in CoAs flux with H$_2$-treated Co powder}
\end{table}

The $g$-factor of the Eu$^{+2}$ spins is obtained from the quantum condition
\bse
\be
g = \frac{h f}{\mu_{\rm B}H_{\rm res}^{\rm int}},
\ee
where $f$ is the microwave frequency.  In our experiments, the X-band microwave frequency was $f=9.390$~GHz.  Thus one obtains
\be
g = \frac{6.709}{H_{\rm res}^{\rm int}\,[{\rm kOe}]}.
\label{Eq:gCalc}
\ee
\ese
Plots of the $g$-factor versus $T$ for Crystals~\#2 and~\#3 obtained from the data in Fig.~\ref{Fig:EuCo2As2c_Hres}(b) using Eq.~(\ref{Eq:gCalc}) are shown in Fig.~\ref{Fig:EuCo2As2c_Hres}(c), where the variation with temperature in Fig.~\ref{Fig:EuCo2As2c_Hres}(b) is inverted.  A discontinuity in the slope of $g$ versus temperature is seen for Crystal~\#3 at $T\approx 65$~K and for Crystal~\#2 at $\approx 90$~K, again suggesting phase transitions at these temperatures in the respective crystals as reflected in the resonance field data in Figs.~\ref{Fig:EuCo2As2c_Hres}(a) and~\ref{Fig:EuCo2As2c_Hres}(b).

Above $\sim 125$~K the $g$~values are close to the expected value of~2.  However, on cooling below $\sim 125$~K, the $g$-factors increase monotonically to values at 50~K that are $\approx 8$\% enhanced above the high-temperature values of $\approx 2.00$.  This low-$T$ enhancement of the $g$-factor is similar to the $\approx 7$\% enhancement of the Eu$^{+2}$ effective moment in \eca\ crystals obtained~\cite{Sangeetha2017} from magnetic susceptibility measurements from 100 to 300~K that was theoretically attributed to spin polarization of the Co $3d$-band electrons by the Eu spins~\cite{Sangeetha2017}.  However, the spatial distribution of this conduction-electron polarization with respect to the Eu spin positions was not determined.  This issue is further discussed in Sec.~\ref{Sec:Summary}.

\subsection{Linewidth}

\begin{figure}
\includegraphics [width=3.3in]{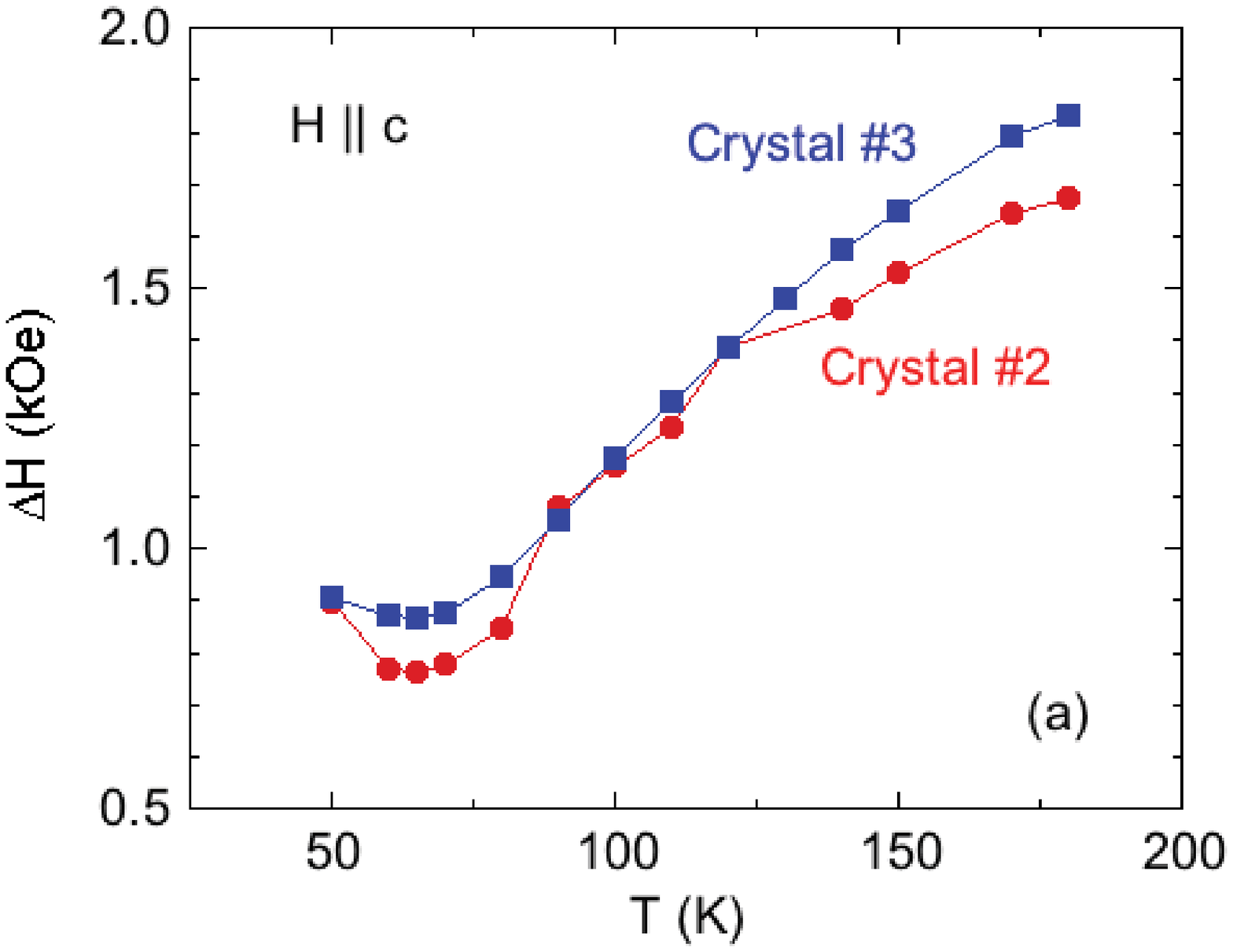}
\includegraphics [width=3.3in]{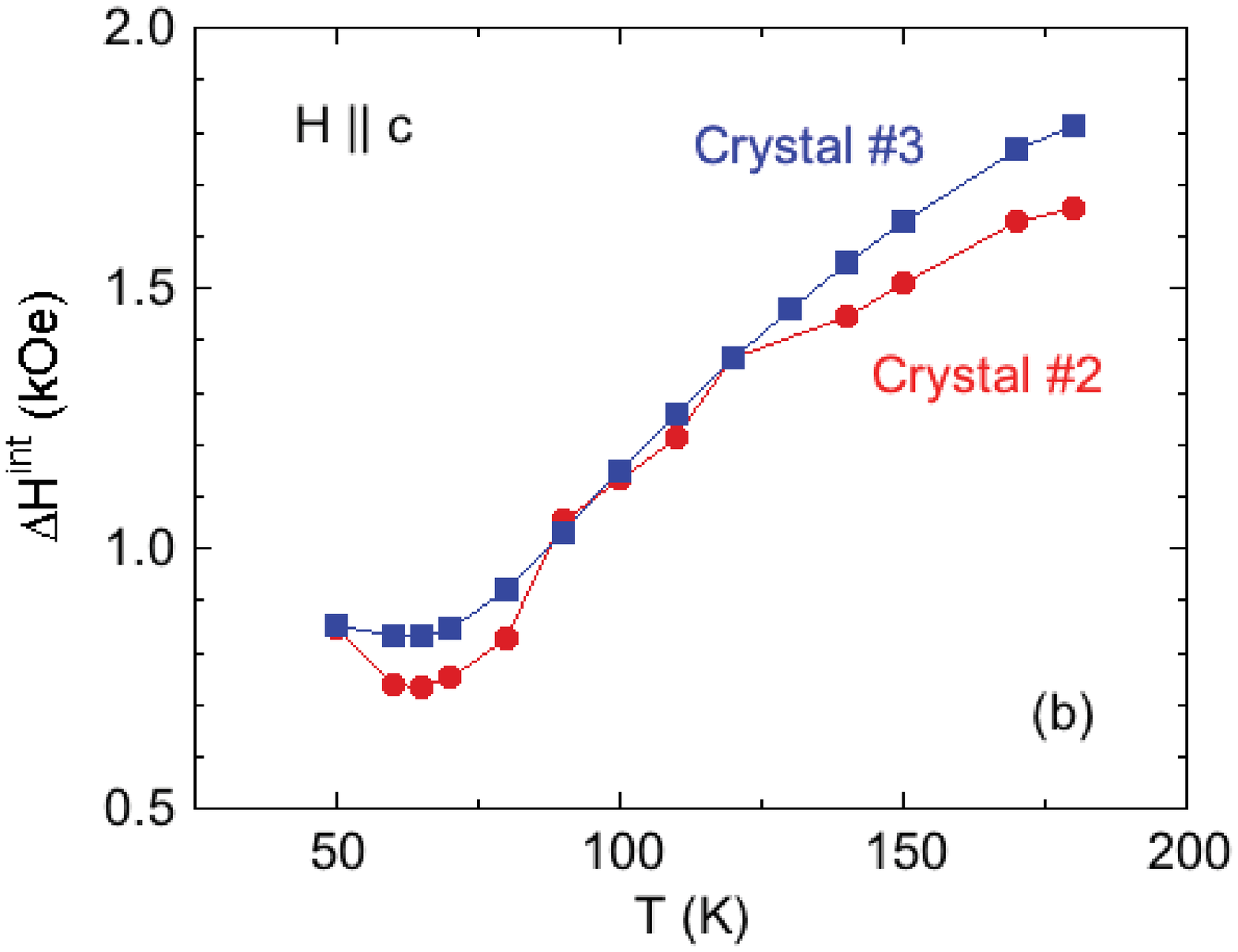}
\includegraphics [width=3.3in]{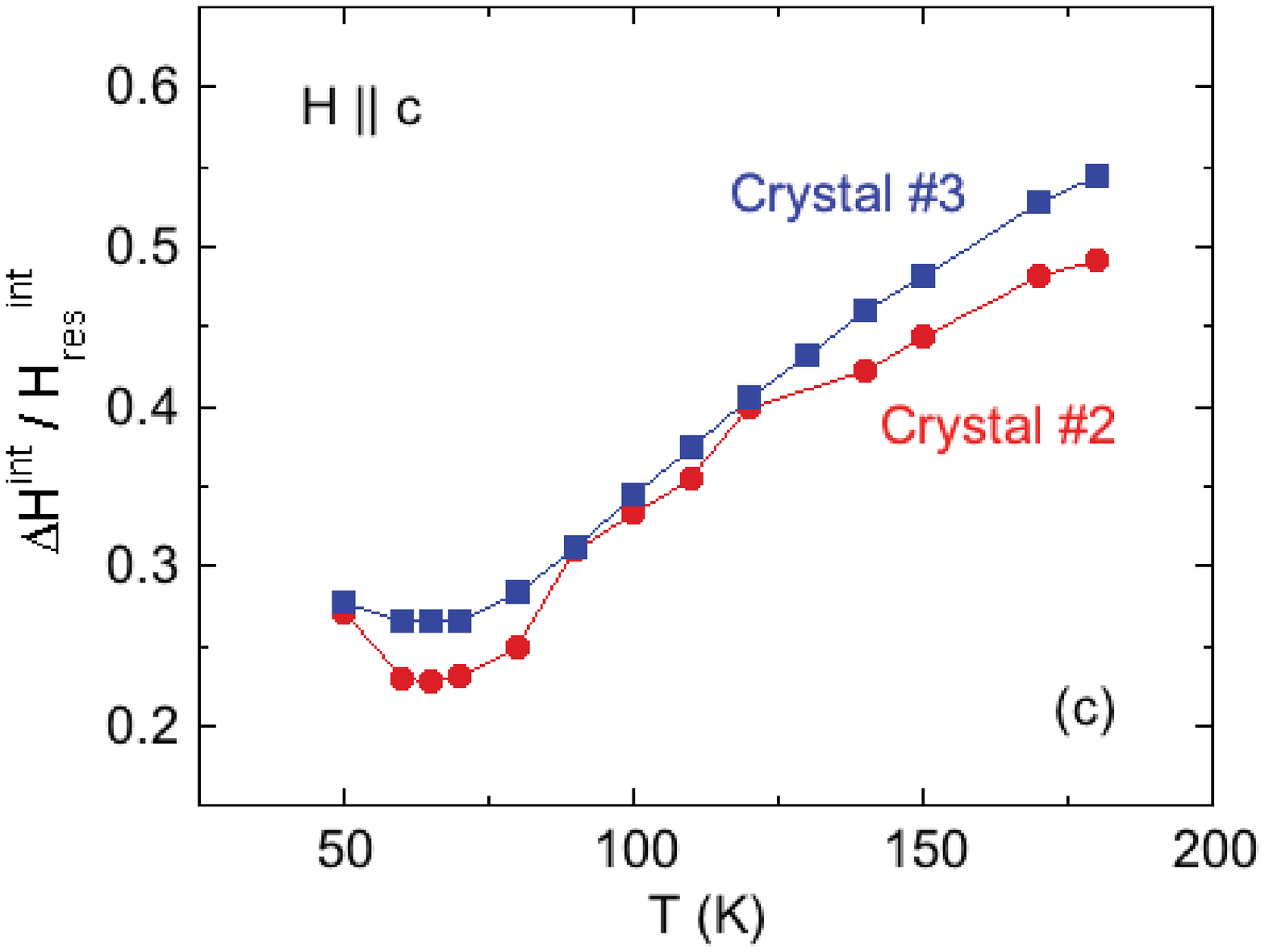}
\caption {(a) Lorentzian half-width $\Delta H$ versus temperature~$T$ for Crystals~\#2 and~\#3 obtained by fitting the field-derivative EPR spectra by Eq.~(\ref{Eq:dsignaldH}).  (b)~Half-width $\Delta H^{\rm int}$ corrected for the demagnetization field versus~$T$ as in Fig.~\ref{Fig:EuCo2As2c_Hres}(b) for $H^{\rm int}$.  (c)~The ratio $\Delta H^{\rm int}/H_{\rm res}^{\rm int}$ versus $T$\@.}
\label{Fig:EuCo2As2c_DeltaH}
\end{figure}

The Lorentzian half-width $\Delta H$ of the resonance versus~$T$ is plotted in Fig.~\ref{Fig:EuCo2As2c_DeltaH}(a) for Crystals~\#2 and~\#3 obtained from fits of the EPR spectra.  The data corrected for the demagnetization field as in Fig.~\ref{Fig:EuCo2As2c_Hres}(b) are shown in Fig.~\ref{Fig:EuCo2As2c_DeltaH}(b).  The latter data for Crystal~\#2 suggest possible phase transitions at $\approx 90$~K and 120~K, whereas the data for Crystal~\#3 do not exhibit clear evidence for any phase transitions.  Overall, the linewidth of both crystals increases with increasing temperature above 70~K as expected for relaxation of the Eu spins by exchange interactions with the conduction electrons.  However the behavior is not linear in~$T$ as expected for such Korringa relaxation.  The average slope between 90~K and 180~K for Crystal~\#2 is 6.7~Oe/K, whereas for Crystal~\#3 the average slope is 8.7~Oe/K\@.  These slope values are in the range found for similar Fe-based ${\rm ThCr_2Si_2}$-structure pnictide compounds containing Eu$^{+2}$ ions~\cite{Ying2010, KrugvonNidda2012}. The ratio $\Delta H^{\rm int}/H_{\rm res}^{\rm int}$ is plotted versus~$T$ in Fig.~\ref{Fig:EuCo2As2c_DeltaH}(c) for the two crystals.  This ratio increases from about 0.25 at $T\approx 60$~K to about 0.5 at $T=180$~K\@.

\section{\label{Sec:Summary} Summary and Discussion}

The theory to fit broad Dysonian EPR spectra for local magnetic moments in metals within the context of the modified Bloch equations was developed.  This included a solution of the absorptive susceptibility  $\chi^{\prime\prime}(\omega)$ at fixed $H$ that is consistent with previous usage.  However, the dispersive susceptibility $\chi^\prime(\omega)$ has a form equivalent to that previously obtained in 1955~\cite{Garstens1955}.  We showed that this form of $\chi^\prime(\omega)$ is valid, since it is derivable via a Kramers-Kronig relation from the expression for $\chi^{\prime\prime}(\omega)$.  The expressions for $\chi^\prime(\omega)$ and $\chi^{\prime\prime}(\omega)$ at fixed field were then converted to $\chi^\prime(H)$ and $\chi^{\prime\prime}(H)$ at fixed~$\omega$ that were later used to fit our field-derivative EPR spectra.

The field derivative of the Dysonian lineshape in Eq.~(\ref{Eq:dsignaldH}) using our expressions for $\chi^\prime(H)$ and $\chi^{\prime\prime}(H)$ in Eqs.~(\ref{Eqs:chi(H)}) were compared with those obtained using the traditional expressions in Eqs.~(\ref{Eqs:ChipppJoshi}).  Rather large differences were found for $\alpha>0$ and for large linewidths, as exemplified in Figs.~\ref{Fig:DysonSpectVSH_DeltH_alpha1} to~\ref{Fig:Field-derivative_lineshapes}.

Excellent fits of the experimental field-derivative CW~EPR spectra for \eca\ by the general Eq.~(\ref{Eq:dsignaldH}) using our Eqs.~(\ref{Eqs:chi(H)}) were obtained in the PM phase at temperatures from 50~K to 180~K for Crystals~\#2 and~\#3.  According to our analysis of the skin depth in comparison to the dimensions of the sample surface perpendicular to the $c$~axis ($\sim$~mm), the expected value of $\alpha$ is unity according to Dyson's theory, in approximate agreeement with our high-temperature data.  However, below $\sim 150$~K $\alpha$ decreased for both crystals.  The reason is not clear, but the decreases may be associated with possible phase transitions at $T < 150$~K in the crystals.

The temperature-dependent EPR data for Crystal~\#3 showed evidence for a second-order phase transition at 65~K, whereas the data for Crystal~\#2 suggested possible transitions at $\sim 90$~K and $\sim 120$~K\@.  The additional EPR signal in Crystal~\#3 that appears at $\sim 1$~kOe in Fig.~\ref{Fig:SNS_Spectra_Fit}(b) at $T \lesssim 70$~K is likely associated with the phase transition in this crystal at $\approx 65$~K rather than with PM impurities.   Additional experiments are required to determine whether these features are associated with phase transitions.

The Lorentzian resonance half-width $\Delta H$ increases monotonically, but nonlinearly, from 70~K to 180~K with an average slope of 6.7 and 8.7~Oe/K for Crystals~\#2 and~\#3, respectively.  These values are in the range found for similar ${\rm ThCr_2Si_2}$-structure Eu$M_2X_2$ compounds, where $M$ = Fe and/or mixtures with other transition metals and $X$ is As and/or mixtures with P\@.

Microscopic information was obtained on the Eu magnetism where the effective moment was reported to be enhanced by about 7\% above the value expected for $g=2$ and $S = 7/2$ from magnetic susceptibility measurements in the PM temperature range from 100 to 300~K~\cite{Sangeetha2017}.  Over the $T$~range $125~{\rm K} \lesssim T\leq 180$~K, the EPR $g$-factor was found to be approximately constant with an unenhanced value $g\approx 2.00$.  On the other hand, on cooling from $\sim 125$~K to 50~K, the $g$-factor continuously increased in both of our crystals by about 8\% to $\approx 2.16$.  The enhancement of the effective moment above 100~K arises from a global FM spin polarization of the Co~$3d$ electrons by the field-aligned Eu spins as theoretically predicted~\cite{Sangeetha2017}.  We speculate that the reason that the EPR $g$-factor is not enhanced in above $\sim 125$~K is that {\it local} short-range FM correlations of the conduction electron with the Eu spins are negligible.  Then on cooling below $\sim 125$~K a crossover occurs wherein the local FM correlations continuously increase with decreasing~$T$, thus enhancing the Eu $g$-factor.

On the other hand, the saturation moments at $H = 140$~kOe obtained in the antiferromagnetically-ordered state at $T=2$~K for crystals from the same growth batches \#2 and~\#3 were $\mu_{\rm sat} = 7.04$ and 7.56~$\mu_{\rm B}$/Eu, respectively, where the first value is hardly enhanced and the second one is enhanced above the value expected for $g=2$ by 8.0\%~\cite{Sangeetha2017}.  The reason there was little enhancement of $\mu_{\rm sat}$ in the crystal from batch \#2 whereas the effective moment of this crystal was enhanced by 8.6\% according to Tables~III and~IV in Ref.~\cite{Sangeetha2017} is unclear.

A similar but opposite dichotomy to that seen for \eca\ growth batch~\#2 was observed for Gd metal containing Gd$^{+3}$ spins-7/2.  Whereas the low-$T$ saturation moment is enhanced from 7.00 to 7.55~$\mu_{\rm B}$/Gd, the effective moment obtained from magnetic susceptibility measurements in the PM state above the ferromagnetic Curie temperature $T_{\rm C} = 294$~K~\cite{Dankov1998} is 7.98(5)~$\mu_{\rm B}$/Gd~\cite{Nigh1963}, which is the same within the errors as predicted for $g=2$.  In two separate studies, EPR measurements of Gd$^{+3}$ ions in Gd metal in the PM state also yielded unenhanced $g$-factor values of 1.95(3) and 1.97, respectively~\cite{Kip1953, Chiba1970}.  It is peculiar that the low-$T$, high-$H$ saturation moment of the Gd$^{+3}$ ions in Gd metal is strongly enhanced whereas the PM effective moment and $g$-factor of the Gd$^{+3}$ ions at temperatures above $T_{\rm C}$ are not.

\acknowledgments

This research was supported by the U.S. Department of Energy, Office of Basic Energy Sciences, Division of Materials Sciences and Engineering.  Ames Laboratory is operated for the U.S. Department of Energy by Iowa State University under Contract No.~DE-AC02-07CH11358.


\end{document}